\pdfoutput=1
\documentclass[12pt]{article}
\usepackage[sort,authoryear]{natbib}
\usepackage{mathtools}
\usepackage{amssymb}
\usepackage{amsthm}
\usepackage{enumitem}
\usepackage[margin=1in]{geometry}
\usepackage{graphicx}
\usepackage{subcaption}
\usepackage[utf8]{inputenc}
\usepackage[english]{babel}
\usepackage{xcolor}
\usepackage[colorlinks = true, citecolor = blue, urlcolor = blue]{hyperref}
\PassOptionsToPackage{hyphens}{url}\usepackage{hyperref}
\usepackage{indentfirst}
\usepackage{apptools}
\usepackage{cleveref}
\usepackage{algorithm}
\usepackage{color, colortbl}
\definecolor{gray}{gray}{0.5}
\definecolor{lightgray}{gray}{0.7}
\usepackage{booktabs}%
\usepackage{orcidlink}
\usepackage{hyperref}
\usepackage[noend]{algpseudocode}

\graphicspath{{fig/}}

\AtAppendix{\counterwithin{lemma}{section}}
\AtAppendix{\counterwithin{proposition}{section}}

\newtheorem{example}{Example} 
\newtheorem{definition}{Definition} 
\newtheorem{proposition}{Proposition}

\title{A Comparison of the Bayesian Posterior Probability and the Frequentist $p$-Value in Testing Equivalence Hypotheses}

\author{Daniel Ochieng \orcidlink{0000-0002-1023-5028} \\ Email; \href{mailto:me@somewhere.com}{owinodan2011@gmail.com} }

\date{\today}

\begin{document}

\maketitle

\rule{\textwidth}{.5pt}
\begin{abstract}
Equivalence tests, otherwise known as parity or similarity tests, are frequently used in ``bioequivalence studies" to establish practical equivalence rather than the usual statistical significant difference. In this article, we propose an equivalence test using both the $p$-value and a Bayesian procedure by computing the posterior probability that the null hypothesis is true. Since these posterior probabilities follow the uniform $[0,1]$ distribution under the null hypothesis, we use them in a Two One-Sided Test (TOST) procedure to perform equivalence tests. For certain specifications of the prior parameters, test based on these posterior probabilities are more powerful and less conservative than those based on the $p$-value. We compare the parameter values that maximize the power functions of tests based on these two measures of evidence when using different equivalence margins. We also derive the correlation coefficient between these two measures of evidence. Furthermore, we also consider the effect of the prior variance on the conservativity and power function of the test based on the posterior probabilities. Finally, we provide examples and a small-scale simulation study to compare their performance in terms of type I error rate control and power in a single test, as well as in multiple testing, considering the power of the false discovery rate procedure.
\end{abstract}

{\bf \href{https://mathscinet.ams.org/mathscinet/msc/msc2020.html}{MSC2020 subject classifications:}} 62F03, 62F15, 62J15. \\
\rule{\textwidth}{1.5pt}

%%%%%%%%%%%%%%%%%%%%%%%%%%%%%%%%%%%%%%%%
\section{Introduction}\label{introduction}
%%%%%%%%%%%%%%%%%%%%%%%%%%%%%%%%%%%%%%%%

%\subsection{Motivation}

Many scientific applications require the comparison of two or more groups based on some parameter of interest. Research in the previous literature has emphasized testing the differences between groups, where, for example, the null hypothesis is
$H: \theta_X=\theta_Y \ \text{versus the alternative} \ K: \theta_X\neq\theta_Y$ for two independent random  variables $X$ and $Y$ where $\theta_X$ and $\theta_Y$ are some parameters of interest like the mean. The problem with such tests is that failure to reject the null hypothesis does not imply equivalence of the two means. A better alternative is to define equivalence tests, for example, using the Two One-Sided Test (TOST) proposed by \cite{schuirmann1987comparison}, which uses the Intersection-Union (IU) test. In this testing procedure, we decompose the null hypothesis into a lower and an upper-sided test, that is, $H_1: \theta\leq \theta_1^*\ \text{and} \ H_2: \theta\geq \theta_2^*,$ where $\theta$ is the parameter of interest and $\theta_1^*$ ($\theta_2^*$) is the lower (upper) equivalence margin. We test the two null hypotheses at level $\alpha\in (0,1)$ without a multiplicity adjustment and in no particular order. We declare equivalence only if we have rejected both $H_1$ and $H_2$.  

Suppose that we have some prior information on the parameters of interest. Incorporating this information into our model-building procedure can lead to more powerful tests (for equal sample sizes) than tests based on some frequentist procedures. The use of Bayesian quantities such as Bayes factors, predictive $p$-values, Bayesian information criterion, and posterior probability as a summary measure of evidence regarding the null hypothesis is increasing due to inherent issues with the $p$-value. Some of these issues, as highlighted by \cite{shi2021reconnecting}, are that the $p$-value does not provide a measure of compatibility of the data with the alternative hypothesis, how to achieve greater power for ordered hypothesis, exaggeration of evidence against the null hypothesis, manipulation of the $p$-value to reach the significance level, and concerns about reproducibility and replicability. Also, the $p$-value, a frequentist measure of evidence, can violate the likelihood principle, or two or more tests can give different $p$-values even though one uses the same likelihood. In some cases, we fail to specify the alternative hypothesis, and Bayesian measures such as posterior probability are reasonable quantities. The posterior probabilities are also easier to compute and are coherent, unlike the frequentist $p$-value.

%%%%%%%%%%%%%%%%%%%%%%%%%%%%%%
\subsection{Related Works}
%%%%%%%%%%%%%%%%%%%%%%%%%%%%%%
The article \cite{Degroot1973DoingWC} reconciles the $p$-value and posterior probability by giving an example in which these two are equal when using an improper prior distribution in a certain class of alternative distributions. The same author, while using a different class of alternative distribution, also shows that the two quantities (the $p$-value and the posterior probability) are closely related to the likelihood ratio.

For a one-sided hypothesis test, 
\cite{casella1987reconciling} provides theoretical results to establish equivalence between the posterior probability and the $p$-value when dealing with continuous symmetric distributions that possess the monotone likelihood ratio property \citep{karlin1956distributions}. The authors considered the infimum of the posterior probability over a broad class of prior distributions as the Bayesian measure of evidence.
These results are further verified in \cite{shi2021reconnecting}. Using a noninformative (also known as uninformative) prior distribution,
\cite{shi2021reconnecting}  establishes an equivalence between the posterior probability and the $p$-value for a one-sided null hypothesis.
 The authors further combine two one-sided tests in opposite directions to establish equivalence between the posterior probability when using a noninformative prior and the $p$-value for a two-sided test. Their approach differs from that of \cite{berger1987testing}, which places a prior probability distribution on the point null.  \cite{berger1987testing} also considers a similar problem for a precise (point or small interval) null hypothesis and shows a conflict between the posterior probability and the $p$-value. The same authors note that the posterior probability and the Bayes factors are much higher than the $p$-value. 
 
 \cite{dasgupta2000correlation} derives the correlation coefficient between the $p$-value and the posterior probability in a one-sided hypothesis test. Of course, as expected, as the variance for the prior distribution increases, the posterior probability tends to the $p$-value both in a one-sided  \citep{dasgupta2000correlation} and the two-sided \citep{basu1992new} hypothesis tests. \cite{zaslavsky2010bayesian} also reconciles the $p$-value with the posterior probability when dealing with a one-sided hypothesis test in discrete distributions. Results also exist proving that the frequentist confidence set can exceed the Bayesian credible set. All the results mentioned previously dealt with a one-dimensional comparison. \cite{Kline2011TheBA} extends this comparison to the case of more than one dimension.
 
 Reconciling Bayesian and frequentist procedures has also been carried out in multiple testing cases. \cite{westfall1997bayesian} shows that the adjusted posterior probability of the null hypothesis is roughly equivalent to the Bonferroni-adjusted $p$-values. \cite{Tadesse2005WaveletTW} extends the work of \cite{westfall1997bayesian} to the case of false discovery rate (FDR) control and gives an application in wavelet thresholding. 

%%%%%%%%%%%%%%%%%%%%%%%%%%%%%%%%%%%%%%%%%%%%%%%%%
\subsection{Summary of Our Contributions}
%%%%%%%%%%%%%%%%%%%%%%%%%%%%%%%%%%%%%%%%%%%%%%%%%
The procedures in the literature for equivalence tests have focused mainly on comparing a single parameter, such as the maximum observed concentration $(C_{max})$ between two groups or even multiple different parameters simultaneously (such as $C_{max}$ and the area under the concentration-time
curve at time $t>0$ ($\mathrm{AUC}_t$)), but still between two groups as in \cite{pallmann2017simultaneous}. In this research, we compare several parameters of the same type with a single control parameter by choosing an equivalence margin and comparing all the parameters to this margin. Therefore, this contributes to the comparison of a single parameter between multiple groups, which has been lacking in the literature. 

Following in the footsteps of \cite{zaslavsky2010bayesian} and \cite{shi2021reconnecting}, we also utilize the connection between the posterior probability of the null hypothesis and the $p$-value for a one-sided test. We use this connection to calculate the posterior probability using a modified $p$-value. We then combine these two one-sided tests and use the result to perform an equivalence test among several parameters. 

Using different prior parameters, the posterior probability for a parameter of interest can be equal to, greater than, or less than the one-sided $p$-value. %We rely on this relationship between the posterior probability and the $p$-value to use the posterior probability in the equivalence test. 
Therefore, depending on the choice of the prior parameters, the posterior probability leads to equally, less, or more powerful tests at constant sample sizes compared to the $p$ value in the TOST procedure. \cite{shi2021reconnecting}  illustrated through a simulation study that as the prior distribution becomes less informative, the equivalence between the posterior probability and the $p$-value becomes more evident. We also illustrate this equality for our $p$-value and the posterior probability using their cumulative distribution functions (CDFs) under the null and alternative hypotheses. 

We are also interested in finding a parameter value $\theta_{max}$ under the alternative hypothesis that gives the maximum power for the test based on the $p$-value and the posterior probability. In the same set-up, we find if the value of this $\theta_{max}$ is affected by the size of the equivalence margin or the values of the prior parameters for a constant equivalence margin. %Furthermore, we investigate the effects of a change in the prior parameters on the value of $\theta_{max}$ when the equivalence margin is constant.

Furthermore, we extend the work of \cite{dasgupta2000correlation} and derive the correlation coefficient between the posterior probability and the  $p$-value for an equivalence test. The correlation coefficient between the $p$-value for a lower-sided and an upper-sided test is also derived. For the sake of completeness, some brief remarks for the two-sided test are also given in the Appendix.

Using simulation studies, we compare the performance of the posterior probability and the $p$-value in terms of type I error control and power under two scenarios. First, when the equivalence margin is constant while the sample size and prior parameters are increasing, and second, when the sample size is constant while the size of the equivalence margin and prior parameters are not. Finally, we also extend our comparison of the two measures of evidence to multiple testing by investigating the power of the false discovery rate procedure.

%%%%%%%%%%%%%%%%%%%%%%%%%%%%%%%%%%%%%%%%%%%%%%%%%
\subsection{Overview of the Rest of the Material}
%%%%%%%%%%%%%%%%%%%%%%%%%%%%%%%%%%%%%%%%%%%%%%%%%

We organize the remaining parts of this article as follows. We describe the procedures used in this research article and the hypothesis considered in Section \ref{equivalence}. In particular, the TOST procedure using the $p$-value is described in Section \ref{frequentist}, while the details of the TOST procedure using the posterior probability are in Section \ref{bayesian}. The properties of the posterior probability compared to the $p$-value for a single test when applied to a discrete distribution are considered in Section \ref{propertiesdiscrete}. 
In particular, we investigate the conservativity, power function, and parameter value that maximize this power function under different scenarios. 
To complement the results in this section, we present in Appendix \ref{appendixB2} a small-scale simulation study comparing the type I error rate control and power for tests based on these two measures of evidence.
In Section \ref{propertiescont}, we extend our comparison to a continuous distribution, but still for a single test. In particular, we consider the effect of prior variance on the conservativity and power function for the test based on the posterior probability. A detailed investigation of the correlation coefficient between these two measures of evidence is also available in this section. 
An extension of this comparison, considering the power of the false discovery rate procedure in multiple testing, is the subject of Section \ref{simulation}. This comparison is for varying equivalence margins, sample sizes, and standard deviation of the prior distribution. Finally, we discuss our results and provide avenues for future research in Section \ref{discussion}.

%%%%%%%%%%%%%%%%%%%%%%%%%%%%%%%%%%%%%
\section{Equivalence Testing}
\label{equivalence}
%%%%%%%%%%%%%%%%%%%%%%%%%%%%%%%%%%%%%

%%%%%%%%%%%%%%%%%%%%%%%%%%%%%%%%%%%%%
\subsection{Introduction}
%%%%%%%%%%%%%%%%%%%%%%%%%%%%%%%%%%%%%
Let us assume that we observe random data $\pmb{X}=(X_1,\ldots,X_n)^{\top}, n \in \mathbb{N}$ where the $X_i, 1\leq i\leq n$ are independent and identically distributed (i.i.d.) random variables following a distribution $\mathbb{P}$ from a parametric family $(\mathbb{P}_{\theta}, \theta\in \Theta\subset \mathbb{R})$. Let $\pmb{\mathcal{X}}$ denote the support point of $\pmb{X}$ and, for simplicity, $f(\pmb{x}|\theta)$ the corresponding density such that $f(\pmb{x}|\theta)\geq 0$ and $\int_{\pmb{\mathcal{X}}}f(\pmb{x}|\theta)d\pmb{x}=1$.
 Furthermore, assume that we have a nonnegative test statistic $T(\pmb{X})$ such that the family of distributions $f(\pmb{x}|\theta)$ has a monotone likelihood ratio (MLR) property in $T(\pmb{X})$.
We are interested in testing the hypothesis that the parameter $\theta$ lies in a fixed interval, where the interval used (equivalence margin) is constant throughout; we can express this hypothesis as 
\begin{equation}
H: \theta \notin (\theta_1^*, \, \theta_2^*)\ \text{~~versus~~} \  K: \theta\in (\theta_1^*, \, \theta_2^*), 
\label{hypothesis}
\end{equation} 
for given numbers $\theta_1^*,\theta_2^* \in \Theta$ such that $\theta_1^*<\theta_2^*$. 

To test several hypotheses $k>1$, we express them as $H_j: \theta_j\notin \Delta$ versus $K_j:\theta_j\in \Delta$ where $\Delta$ is the range of values in the interval between $\theta_1^*$ and $\theta_2^*$, which shall remain constant for $j\in \{1,\ldots,k\}$ and $k$ is the multiplicity of the problem. Since the posterior probabilities for a one-sided test problem follow the uniform $[0,1]$ distribution under the null hypothesis \citep{labbe2007multiple}, we can use them instead of the $p$-values in any multiple testing procedure based on the $p$-value. 
In this section, we restrict our description to the case $k=1$ and highlight how we can use the $p$-value and the posterior probability to test this hypothesis. We consider the case of multiple hypothesis tests in Section \ref{simulation}. We use $t\in (0,1)$ and $\alpha\in (0,1)$ interchangeably to denote the significance level.  

%%%%%%%%%%%%%%%%%%%%%%%%%%%%%%%%%%%%%
\subsection{The Frequentist Approach}
\label{frequentist}
%%%%%%%%%%%%%%%%%%%%%%%%%%%%%%%%%%%%%
In this section, we describe how the hypothesis in Equation \eqref{hypothesis} can be tested using the Two One-Sided Test (TOST) procedure based on the least favorable configuration (LFC) $p$-value. When using the TOST procedure, we decompose the equivalence hypothesis into a lower- and an upper-tailed test as mentioned earlier. Therefore, for the upper (right)-tailed test
\begin{equation}
H_{r}:\theta\leq \theta_1^* \text{~~versus~~}  K_{r}:\theta> \theta_1^*,
\label{upper_tailed}
\end{equation}
the $p$-value is
$P_r(\pmb{X})=\mathbb{P}_{\theta_1^*}(T(\pmb{X})\geq  C_n)$, where $\theta_1^*$ is a prespecified constant, $C_n=F^{-1}_{\theta_1^*}(1-\alpha)$
is the critical constant, and $F^{-1}(\cdot)$ denotes the quantile of a random variable.
%$F^{-1}(\bullet)$ is the inverse of a cumulative distribution function, and $\alpha\in (0,1)$ is the significance level. 

Similarly, for the lower (left)-tailed test 
\begin{equation}
H_{l}:\theta\geq \theta_2^* \text{~~versus~~} K_{l}:\theta< \theta_2^*,
\label{lower_tailed}
\end{equation}
the $p$-value is
$P_l(\pmb{X})=\mathbb{P}_{\theta_2^*}(T(\pmb{X})\leq D_n),$ where again $\theta_2^*$ is a prespecified constant and $D_n=F^{-1}_{\theta_2^*}(\alpha)$ is the critical constant. 
%In the usual TOST procedure, one performs both the lower-tailed and upper-tailed tests at level $\alpha$. One then rejects the equivalence hypothesis if and only if one rejects both the null hypotheses in Equations \eqref{upper_tailed} and \eqref{lower_tailed}. 
In the TOST procedure, as explained earlier, we perform each test in \eqref{upper_tailed} and \eqref{lower_tailed} at level $\alpha$, and rejecting both tests leads to rejection of the equivalence hypothesis. 
Next, we give a proposition stating that the $p$-value for the equivalence problem is the maximum of the $p$-values for the lower-tailed and upper-tailed tests.

\begin{proposition}
For a fixed but arbitrary significance level $\alpha\in(0,1)$ and the true parameter under the null hypothesis $\theta_1^*$ or $\theta_2^*$, we reject the null hypothesis for the hypothesis in Equation \eqref{hypothesis} if the maximum of the $p$-values for the lower- and upper-tailed tests is such that $P_{f}(\pmb{X})\leq \alpha$ where
$  
P_{f}(\pmb{X})=\mathrm{max}\{P_{r}(\pmb{X}),P_{l}(\pmb{X})\}=\mathbb{P}_{\theta_1^*}(C_n\leq T(\pmb{X})\leq D_n)=\mathbb{P}_{\theta_2^*}(C_n\leq T(\pmb{X})\leq D_n),  
$ where the last equality holds only when $\theta_1^*$ and $\theta_2^*$ are symmetric.
\label{ch4:proposition1}
\end{proposition}
The proof for Proposition \ref{ch4:proposition1} is similar to that for Lemma $1$ in \cite{ochieng2024bmultiple}. Next, we present the Bayesian approach, which proceeds similarly, except that we are now using posterior probabilities.
%Denote the resulting $k$ $p$-values by $p_1,\ldots,p_k$.

%%%%%%%%%%%%%%%%%%%%%%%%%%%%%%%%%%%%%
\subsection{The Bayesian Approach}
\label{bayesian}
%%%%%%%%%%%%%%%%%%%%%%%%%%%%%%%%%%%%%

Let $\pi(\theta)>0$ for all $\theta$, such that $\int_{\Theta}\pi(\theta)d\theta=1$, be a prior distribution for the unknown parameter $\theta\in \Theta$ and $f(\pmb{x}|\theta)$ the probability distribution of $\pmb{x}$ given $\theta$ as defined earlier. The posterior distribution is then given by $\pi(\theta|\pmb{x})=f(\pmb{x}|\theta)\pi(\theta)/m_\pi(\pmb{x})$ where $\pi(\theta|\pmb{x})\geq 0$, $\int_{\Theta}\pi(\theta|\pmb{x})d\theta=1$, and $m_\pi(\pmb{x})$ 
is the marginal probability distribution of $\pmb{X}$ under the prior density $\pi(\theta)$. The marginal distribution is given by
$m_\pi(\pmb{x})=\int_{\Theta}f(\pmb{x}|\theta)\pi(\theta)d\theta$ for a continuous distribution and $m_\pi(\pmb{x})=\sum_\Theta f(\pmb{x}|\theta)\pi(\theta)$ for a discrete distribution. 

We again decompose the hypothesis in Equation \eqref{hypothesis} into a lower-and upper-sided test and perform each test at the level $\alpha\in (0,1)$. For the upper-tailed test in Equation \eqref{upper_tailed}, we have the posterior probability as
\begin{equation}
P_r(\theta\leq \theta_1^*|\pmb{x},n)=\int_{-\infty}^{\theta_1^*}\pi(\theta|\pmb{x})d\theta.
\label{posterior1}
\end{equation}
We reject the null hypothesis at the significance level $\alpha$ if $P_r(\theta\leq \theta_1^*|\pmb{x},n)\leq \alpha$. Similarly, for the lower-tailed test in Equation \eqref{lower_tailed}, we have the posterior probability as
\begin{equation}
P_l(\theta\geq \theta_2^*|\pmb{x},n)=\int_{\theta_2^*}^{\infty}\pi(\theta|\pmb{x})d\theta.
\label{posterior2}
\end{equation}
Again, we reject the null hypothesis at the significance level $\alpha$ if $P_l(\theta\geq \theta_2^*|\pmb{x},n)\leq \alpha$. For the equivalence hypothesis, we summarize our results in Proposition \ref{ch4:lemma2}.
\begin{proposition}
\label{proposition_posterior_probs}
For a fixed but arbitrary significance level $\alpha\in(0,1)$ and a chosen true parameter under the null hypothesis $\theta_1^*$ or $\theta_2^*$, we reject the null hypothesis in Equation \eqref{hypothesis} if the overall posterior probability is such that $P_{b}(H|\pmb{x})\leq \alpha$ where
\begin{align*}
P_{b}(H|\pmb{x})&=P_l(\theta\geq \theta_2^*|\pmb{x},n)+P_r(\theta\leq \theta_1^*|\pmb{x},n),\\
&=1-P(\theta_1^*<\theta<\theta_2^*|\pmb{x},n).
\label{p_overall}
\end{align*}
\label{ch4:lemma2}
\end{proposition}
%We now give numerical examples to compare some of the properties of these two measures of evidence.
In the next section, we consider numerical examples to compare some of the properties of these two measures of evidence.
%The proof of Proposition \ref{ch4:lemma2} is straightforward.

%%%%%%%%%%%%%%%%%%%%%%%%%%%%%%%%%%%%%%%%%%
\section{Application to Discrete Distributions}\label{propertiesdiscrete}
%%%%%%%%%%%%%%%%%%%%%%%%%%%%%%%%%%%%%%%%%%

%%%%%%%%%%%%%%%%%%%%%%%%%%%%
\subsection{Introduction}
%%%%%%%%%%%%%%%%%%%%%%%%%%%%

In this section, we are interested in investigating some of the properties of our posterior probability in comparison to the $p$-value. These properties include the conservativity and behavior of the power function in different scenarios. We investigate them by plotting the cumulative distribution functions for the two measures of evidence. We also investigate the parameter value that maximizes the power of these two quantities and their relation to the size of the equivalence margin and the prior parameters. We first consider an example that we will use throughout this section to investigate these properties.

\begin{example}[Binomial distribution]\label{ch4:binomial_model}
Assume that our data $X_i$, $1 \leq i \leq n$ are i.i.d. Bernoulli variables with parameter $\theta\in (0,1)$, $Bernouli(\theta)$ for short. 
A sufficient test statistic is $T(\pmb{X})=\sum_{i=1}^n X_i$, which follows a binomial distribution with parameters $n$ and $\theta$, and we denote this by $Bin(n,\theta)$. 
We also assume that we have the conjugate Beta prior $\pi(\theta)=\mathrm{Beta}(p,q)$ for the unknown binomial parameter $\theta\in (0,1)$ where $p$ and $q$ are the parameters of the prior distribution. 
Then, the posterior distribution $\pi(\theta|\pmb{x})$ is also a Beta distribution, that is, $\mathrm{Beta} (p+s,q+n-s)$, where $s=\sum_{i=1}^n x_i$. 
Let $x_m=p+s$ and $n_m=n+p+q-1$, then for the upper-tailed test in Equation \eqref{upper_tailed}, we have
\begin{equation*}
P_r(\theta\leq \theta_1^*|\pmb{x},n)=1-\sum_{y=0}^{x_m-1}\binom{n_m}{y}(\theta_1^*)^y (1-\theta_1^*)^{n_m-y},
\label{eq:01}
\end{equation*} 
using the posterior probability calculation in Equation \eqref{posterior1} and the relationship between Binomial and Beta random variables. Similarly, for the lower-tailed test in Equation \eqref{lower_tailed}, we have
\begin{equation*}
P_l(\theta\geq \theta_2^*|\pmb{x},n)=\sum_{y=0}^{x_m-1} \binom{n_m}{y}(\theta_2^*)^y (1-\theta_2^*)^{n_m-y},
\end{equation*}
using the posterior probability calculation in Equation \eqref{posterior2}. 
We use these two probabilities to calculate the overall posterior probability in Proposition \ref{proposition_posterior_probs}. To calculate the $p$-value, the critical constants $C_n$ and $D_n$ are given by
$C_n=F^{-1}_{Bin(n, \theta_1^*)}(1-t)$ and $D_n=F^{-1}_{Bin(n, \theta_2^*)}(t)$ for $t\in (0,1)$ where $F^{-1}(\cdot)$ denotes the quantile of a binomial random variable with parameters $n$ and $\theta.$ With these two quantities determined, the $p$-value is calculated as explained in Proposition \ref{ch4:proposition1}.
\label{ex:01}
%In the rest of the material, the subscripts for the prior distribution for the $i^{th}$ parameter are dropped and denote it as $\mathrm{Beta}(p,q)$. 
\end{example}

%The results in Example \ref{ex:01} hold for any symmetric beta prior distribution that is, $Beta(a,b)$ with $a=b$. As the values of $a,b\to 0$, the posterior probability tends to the $P$-value. In general, the  posterior probability is always less than the LFC-based $P$-value when no adjustment is made to the observed value and the sample size. Exact equality can however be obtained in the case of large sample sizes. 

%%%%%%%%%%%%%%%%%%%%%%%
\subsection{Conservativity and Power}
\begin{figure}[ht!]
\begin{center}
\includegraphics[width=16cm,height=8cm]{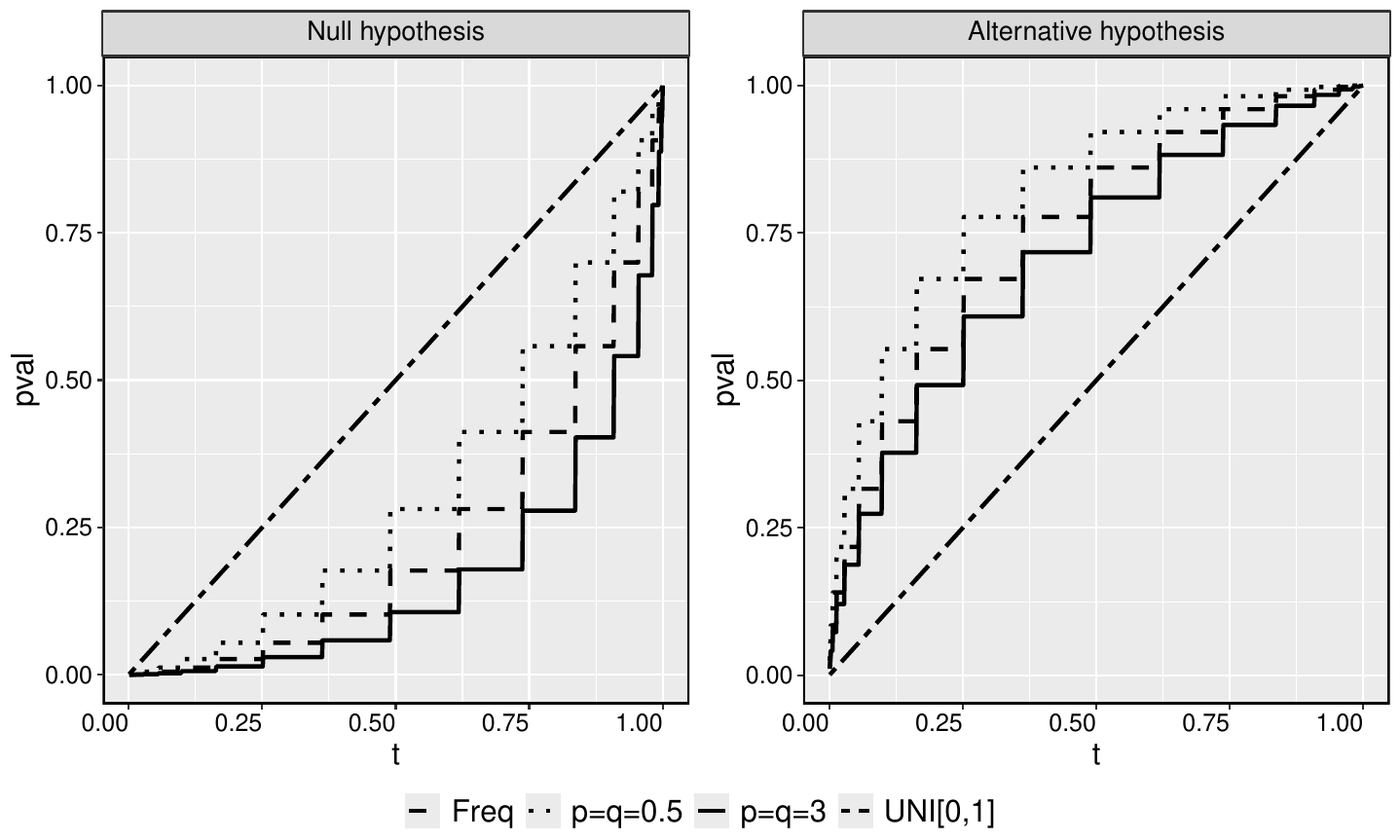}
\caption{An illustration of the conservativeness and power of the test based on the $p$-value and the posterior probability using a sample of size $n=50$, equivalence margin $(\theta_1^*,\theta_2^*)=(0.25,0.75)$,  and with different significance levels $t\in (0,1)$. We also use two different sets of prior distributions: $\mathrm{Beta}(0.5,0.5)$ denoted by $p=q=0.5$ and $\mathrm{Beta}(3,3)$ denoted by $p=q=3$.}
\label{Figure_05}
\end{center}
\end{figure}

%%%%%%%%%%%%%%%%%%%%%%%
We plot Figure \ref{Figure_05} to compare the conservativity and the power functions for the test based on the posterior probability and the $p$-value. To generate Figure \ref{Figure_05}, we use
a sample of size $n=50$, the equivalence margin $(\theta_1^*,\theta_2^*)=(0.25,0.75)$, two different sets of prior parameters $\mathrm{Beta}(0.5,0.5)$ and $\mathrm{Beta}(3,3)$, and with different significance levels $t\in(0,1)$.

In Figure \ref{Figure_05}, when we use the prior distribution $\mathrm{Beta}(0.5,0.5)$, the posterior probability is less conservative compared to the $p$-value. As the prior parameters increase, so that the prior variance $\tau^2=(pq)/(p+q)^2(p+q+1)$ also increases, the conservativity of the posterior probability increases. For example, for $\mathrm{Beta}(3,3)$, the posterior probability is now more conservative than the $p$-value. The power functions have an opposite trend so that for large values of the prior parameters (again implying a larger prior variance $\tau^2$), for example $\mathrm{Beta}(3,3)$, the test based on the posterior probability has less power than the one based on the $p$-value. For small prior parameters such as $\mathrm{Beta}(0.5,0.5)$, the test based on posterior probability has more power than that based on the $p$-value.

%%%%%%%%%%%%%%%%%%%%%%%%%%%%%%%%%
\subsection{Maximum Power}
\begin{figure}[ht!]
\begin{center}
\includegraphics[width=16cm,height=10cm]{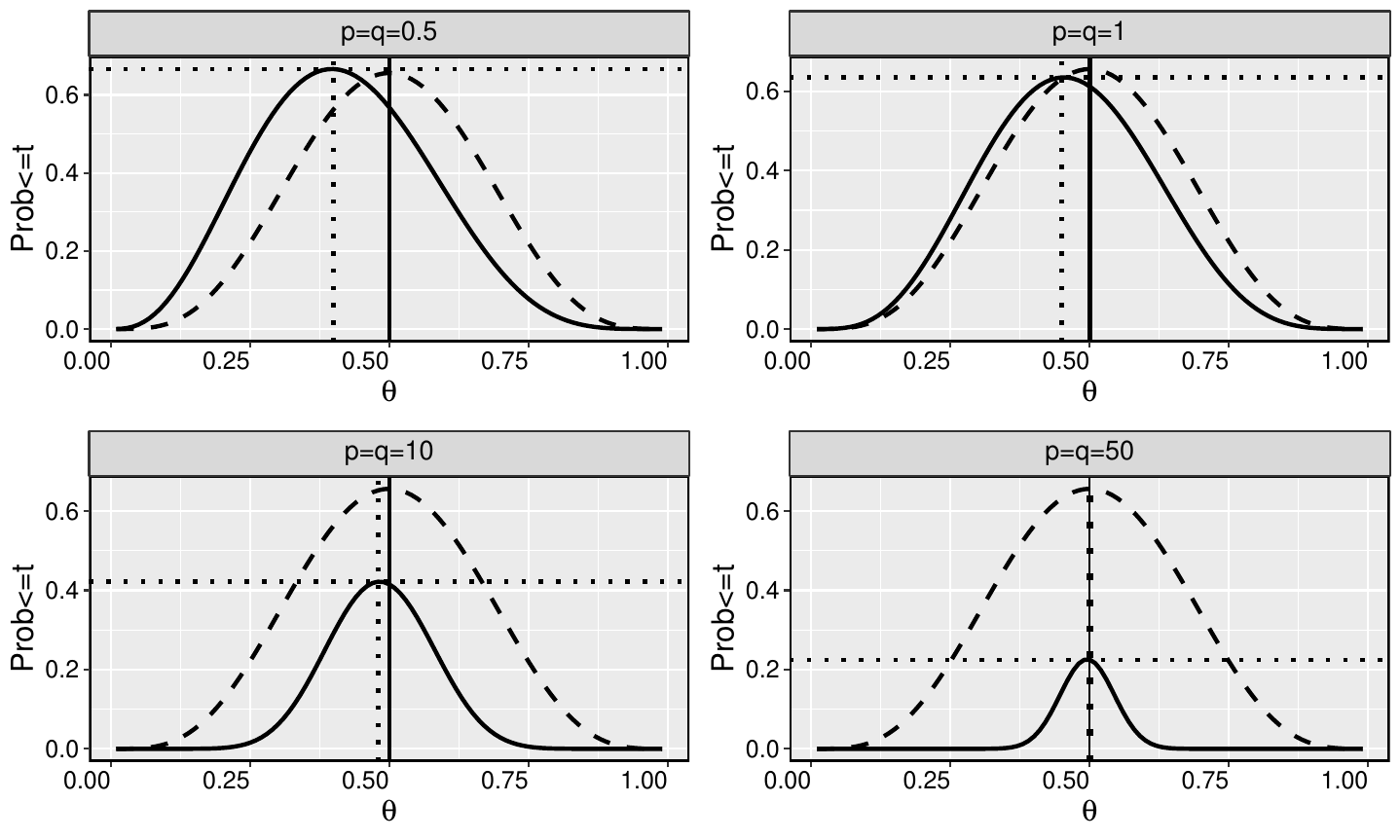}
\caption{An illustration of the power function for the test based on the $p$-value and the posterior probability for a sample of size $n=10$ and equivalence margin $(\theta_1^*,\theta_2^*)=(0.2,0.8)$. We use the Beta prior distribution with different parameters: $\mathrm{Beta}(0.5,0.5)$, $\mathrm{Beta}(1,1)$, $\mathrm{Beta}(10,10)$, and $\mathrm{Beta}(50,50)$ in the four panels, respectively, from left to right. We also use different true parameters $\theta\in (0,1)$ and note that the continuous curve is for the posterior probability while the dashed curve is for the $p$-value. The vertical lines intersect the power curves at their maximum points (dotted line for the posterior probability and continuous line for the $p$-value). The horizontal line illustrates the decrease in power for the posterior probability.}
\label{Figure_max_power}
\end{center}
\end{figure}

%%%%%%%%%%%%%%%%%%%%%%%%%%%%%%%%%
In this section, we are interested in finding a parameter value $\theta_{max}$ in the alternative hypothesis that gives the maximum power for the test based on the $p$-value and the posterior probability. We are also interested in finding how this parameter relates to the prior parameters for the posterior probability and if it is affected by the size of the equivalence margin. 
Choosing such a parameter value ensures that the test based on the $p$-value or the posterior probability achieves the maximum power. To answer these questions, we generate Figure \ref{Figure_max_power} where we have used a sample of size $n=10$ and set the equivalence margin at $(\theta_1^*,\theta_2^*)=(0.2,0.8)$. We chose different prior parameters for the Beta distribution: $\mathrm{Beta}(0.5,0.5)$, $\mathrm{Beta}(1,1)$, $\mathrm{Beta}(10,10)$, and $\mathrm{Beta}(50,50)$ in the four panels, respectively, from left to right. We also use different true parameters $\theta$ in the parameter space $(0,1)$.

We realize from Figure \ref{Figure_max_power} that the true parameter that maximizes the power function always occurs at the value $\theta=0.5$ for the $p$-value. This observation remains the same for any size of the equivalence limit. For the posterior probability, this parameter $\theta_{max}$ is less than $0.5$ and only moves closer to $0.5$ as the prior parameter increases. To achieve exact equality of $\theta_{max}=0.5$ with these small sample sizes, one has to use larger values of the prior parameter, for example $\mathrm{Beta}(50,50)$, as illustrated in the last panel. Unfortunately, at these high prior parameters, there is a loss in power. This trend remains unchanged for all sizes of the equivalence margin. It also remains the same for larger sample sizes, the only difference being a slight loss in power for large values of the prior parameters. The TOST procedure, as expected, has symmetric and unimodal power curves. This unimodality also originates from the fact that the distribution under investigation belongs to an exponential family, which has the MLR property. 

\begin{figure}[ht!]
\begin{center}
\includegraphics[width=16cm,height=10cm]{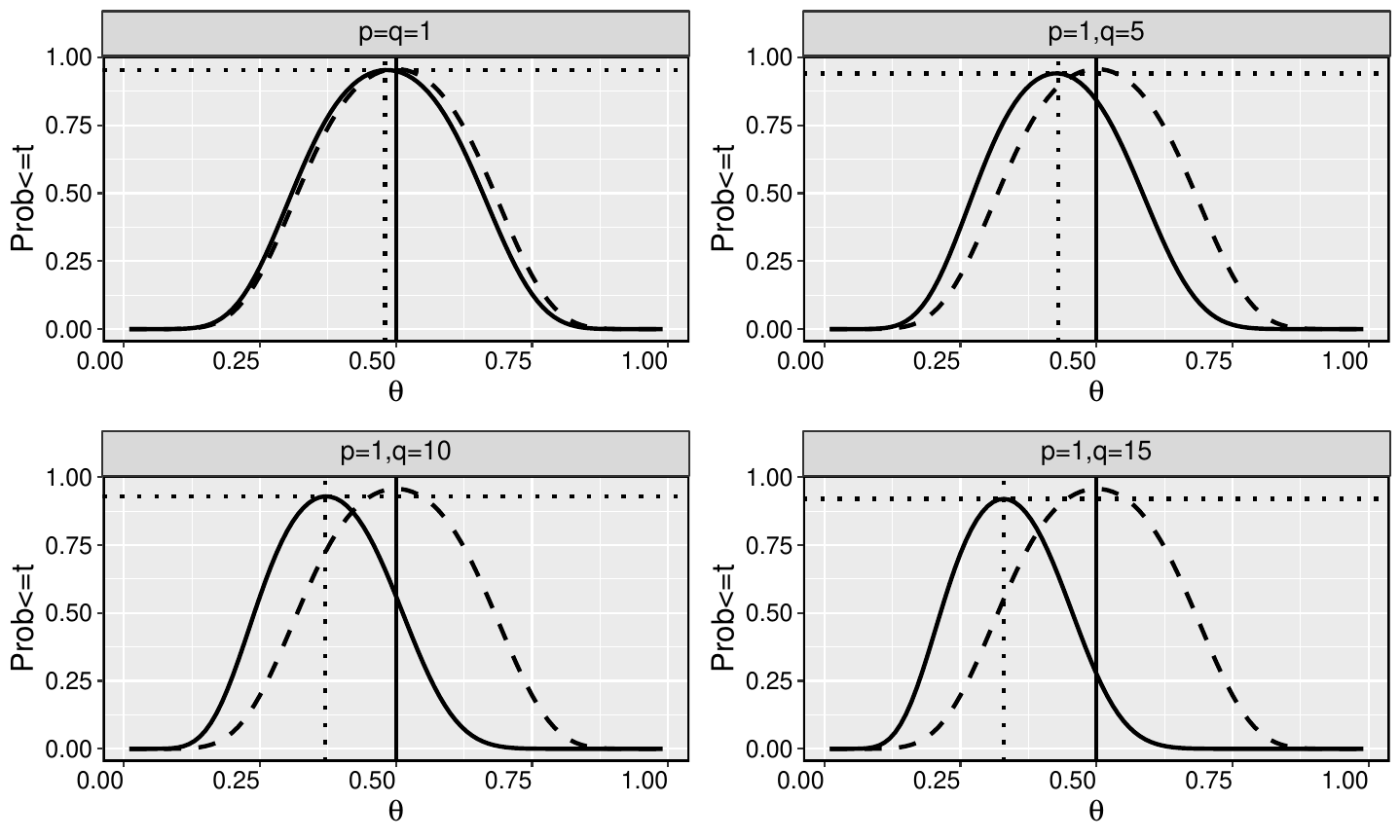}
\caption{An illustration of the power function for the test based on the $p$-value and the posterior probability for a sample of size $n=30$ and equivalence margin $(\theta_1^*,\theta_2^*)=(0.2,0.8)$. We use the Beta prior distribution with different parameters: $\mathrm{Beta}(1,1),\mathrm{Beta}(1,5),\mathrm{Beta}(1,10)$, and $\mathrm{Beta}(1,15)$ in the four panels, respectively, from left to right. We also use different true parameters $\theta\in (0,1)$. The continuous curve is for the posterior probability while the dashed curve is for the $p$-value. The vertical lines intersect the power curves at their maximum points (dotted line for the posterior probability and continuous line for the $p$-value). The horizontal line illustrates the decrease in power for the posterior probability.}
\label{Figure_changing_prior_b}
\end{center}
\end{figure}

We also investigate the effect on the parameter that maximizes the power function when using prior parameters that are not symmetric.
To do this, we maintain all the other configurations as in Figure \ref{Figure_max_power} and only change the sample size to $n=30$ and the priors to $\mathrm{Beta}(1,1),\mathrm{Beta}(1,5),\mathrm{Beta}(1,10)$, and $\mathrm{Beta}(1,15)$ for the four panels, respectively, from left to right. The resulting plot is shown in Figure \ref{Figure_changing_prior_b}.  

We find in Figure \ref{Figure_changing_prior_b} that for a $\mathrm{Beta}(p,q)$ prior and as the value of the prior parameter $q$ increases, while $p$ is kept constant, the power curve for the posterior probability shifts to the left of that for the $p$-value. The maximum of this power curve also moves away from the one for the $p$-value when we change the parameters in this manner. This change also led to a slight drop in power, but not as much as in the last panel of Figure \ref{Figure_max_power}. Reversing this change so that now the prior parameter $q$ is kept constant while the parameter $p$ is increasing, then the reverse happens so that the maximum parameter and the power curve for the posterior probability will now move to the right of the one for the $p$-value. There is also a slight decrease in power similar to that in Figure \ref{Figure_changing_prior_b}. 

\begin{figure}[ht!]
\begin{center}
\includegraphics[width=12cm,height=10cm]{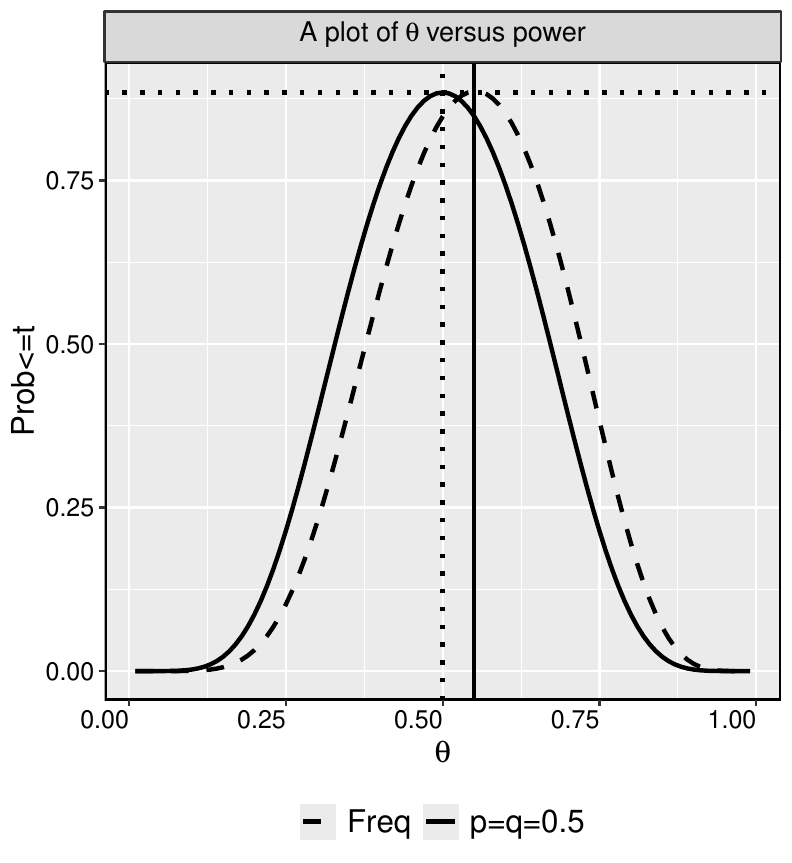}
\caption{An illustration of the power function for the test based on the $p$-value and the posterior probability for a sample of size $n=20$ and equivalence margin $(\theta_1^*,\theta_2^*)=(0.2,0.8)$. We use the $\mathrm{Beta}(0.5,0.5)$ prior distribution, different true parameters $\theta\in (0,1)$ and significance levels $\alpha_1=0.025$ 
for the upper-tailed test and $\alpha_2=0.1$ for the lower-tailed test. The continuous curve is for the posterior probability while the dashed curve is for the $p$-value. The vertical lines intersect the power curves at their maximum points (dotted line for the posterior probability and continuous line for the $p$-value). The horizontal line illustrates the value of the power for the posterior probability.}
\label{Figure_max_power_alpha}
\end{center}
\end{figure}

Finally, we consider the effect of using different significance levels $\alpha_1\neq \alpha_2$ in the TOST procedure. We illustrate this in Figure \ref{Figure_max_power_alpha} using
a sample of size $n=20$, equivalence margin $(\theta_1^*,\theta_2^*)=(0.2,0.8)$, prior distribution $\mathrm{Beta}(0.5,0.5)$, different true parameters $\theta\in (0,1)$, and significance levels $\alpha_1=0.025$ for the upper-tailed test and $\alpha_2=0.1$ for the lower-tailed test. The resulting plot is shown in Figure \ref{Figure_max_power_alpha}.

%\begin{figure}
%\centering
%\scalebox{1}{
%\hspace{-0.025 \textwidth}
%\begin{minipage}{0.35\textwidth}
%\centering
%\includegraphics[width=\textwidth]{fig/class_dist_to_winner.pdf}
%\end{minipage}
%\hspace{0.05 \textwidth}
%\begin{minipage}{0.35\textwidth}
%\centering
%\includegraphics[width=\textwidth]{Figure_04_max_power_alpha.pdf}
%\end{minipage}
%}
%\caption{.}
 %   \label{Figure_max_power_alpha}
%\end{figure}

In Figure \ref{Figure_max_power_alpha} for unequal significance levels and small symmetric prior parameters, the parameter value that maximizes the power of the test based on the posterior probability is always $\theta_{max}=0.5$. This observation is no longer valid for the $p$-value, and the wider the difference between the significance levels $\alpha_1$ and $\alpha_2$, the more $\theta_{max}$ moves away from $0.5$.

%%%%%%%%%%%%%%%%%%%%%%%%%%%%%%%%%%%%%%%%%%
\section{Application to Continuous Distributions}\label{propertiescont}
%%%%%%%%%%%%%%%%%%%%%%%%%%%%%%%%%%%%%%%%%%

%%%%%%%%%%%%%%%%%%%%%%%%%%%%
\subsection{Introduction}
%%%%%%%%%%%%%%%%%%%%%%%%%%%%
We now switch our attention from the discrete case to the case of continuously distributed random variables. We investigate the effect of noise on the cumulative distribution function (CDF) of the $p$-value under the null hypothesis and the alternative hypothesis in Section \ref{effectofnoise}.
% \textcolor{red}{We illustrate this for the $p$-value only since the same trend occurs for the posterior probability.}
We investigate this effect by varying the value of the standard deviation in the CDF of the $p$-value. We are also interested in the correlation coefficient between these two measures of evidence (the $p$-value and the posterior probability). This investigation is the subject of Section \ref{effectofcorrelation}. The goal here is to check whether the size of the equivalence margin affects this correlation coefficient. Furthermore, we derive the correlation coefficient between the $p$-values from the partial tests in Equations \eqref{upper_tailed} and \eqref{lower_tailed} in the same section. First, we state the following general definition for the $p$-value and its CDF in the continuous case. 

\begin{definition}
Let $X_1,\ldots, X_n$ be independent and identically distributed random variables, and we are interested in testing the hypothesis in Equation \eqref{hypothesis}. Let $T(\pmb{X})$ be a sufficient test statistic. A test procedure $\psi(\pmb{X})$ that leads to a rejection of the null hypothesis when $C_n\leq T(\pmb{X})\leq D_n$ is assumed to be available where $C_n, D_n\in \mathbb{R}$, such that $C_n<D_n$ are the critical constants. 
To find the value of these constants, one solves the equations $\mathbb{E}_{\theta_1^*}[\psi(\pmb{X})]=\mathbb{E}_{\theta_2^*}[\psi(\pmb{X})]=\alpha$
%\[\mathbb{E}_{\theta_1^*}[\psi(\pmb{X})]=\mathbb{P}_{\theta_1^*}(C_n\leq T(\pmb{X})\leq D_n)=\alpha,\] and  \[\mathbb{E}_{\theta_2^*}[\psi(\pmb{X})]=\mathbb{P}_{\theta_2^*}(C_n\leq T(\pmb{X})\leq D_n)=\alpha,\] 
simultaneously where $\mathbb{E}(\cdot)$ denotes the mathematical expectation of a random variable. For symmetric distributions, we determine the critical constants by utilizing the symmetry of those distributions. For nonsymmetric distributions, we determine the constants using a numerical solution, like the Newton-Raphson algorithm. The $p$-value for this test procedure from  Proposition \ref{ch4:proposition1} is thus defined as 
\begin{equation}
P_f(\pmb{X})=\mathbb{P}_{\theta_1^*}(C_n\leq T(\pmb{X})\leq D_n)=\mathbb{P}_{\theta_2^*}(C_n\leq T(\pmb{X})\leq D_n).
\label{normal_pval}
\end{equation}
The cumulative distribution function for this $p$-value is given by
\begin{equation}
\mathbb{P}_{\theta}(P_f(\pmb{X})\leq t)=\mathbb{P}_{\theta}(C_n\leq T(\pmb{X})\leq D_n),
\label{normal_pval_cdf}
\end{equation}
where $\theta_1^*<\theta<\theta_2^*.$
\end{definition} 
We now give an example when dealing with normally distributed random variables to illustrate how to calculate the $p$-value $P_f(\pmb{X})$ and its CDF in practice. 
%We then plot Figure \ref{Figure_05_equivalence_normal} to illustrate the effect of different levels of noise on the conservativity and power function of the test procedure based on this $p$-value.

\begin{example}[Testing the Mean of the Normal Distribution]
\label{example_normal}
Let $X_1,\ldots, X_n$ be i.i.d. random variables from the normal distribution with unknown mean $\theta \in \mathbb{R}$ and known variance $\sigma^2\in\mathbb{R}^+$, and we are interested in testing the hypothesis in \eqref{hypothesis}. The sufficient test statistic in this case is $T(\pmb{X})=\sum_{i=1}^nX_i$. To solve for the critical constants $C_n$ and $D_n$, assume that $C_n+D_n=n(\theta_1^*+\theta_2^*)$ and set $C_n=C_0$ where $C_0$ is an arbitrary constant. Then we have $D_n=n(\theta_1^*+\theta_2^*)-C_0$ so that the two critical constants require solving the equation
\[\Phi\bigg(\dfrac{n\theta_2^*-C_0}{\sigma \sqrt{n}}\bigg)-\Phi\bigg(\dfrac{C_0-n\theta_1^*}{\sigma \sqrt{n}}\bigg)=\alpha,\]
where $\Phi(\cdot)$ denotes the CDF of a standard normal random variable. 
This equation can then be solved using normal tables. These critical constants can also be approximated by $C_n=\Phi_{(n\theta_1^*,\sigma\sqrt{n})}^{-1}(1-t)$ and $D_n=\Phi_{(n\theta_2^*,\sigma\sqrt{n})}^{-1}(t)$, respectively, where $\Phi_{(\theta,\sigma)}^{-1}$ denotes the quantile of a normal random variable with mean $\theta$ and variance $\sigma^2$.
%\[\Phi\bigg(\dfrac{n\theta_2^*-C_0}{\sigma \sqrt{n}}\bigg)-\Phi\bigg(\dfrac{C_0-n\theta_1^*}{\sigma \sqrt{n}}\bigg)=\Phi\bigg(\dfrac{n\theta_1^*-C_0}{\sigma \sqrt{n}}\bigg)-\Phi\bigg(\dfrac{C_0-n\theta_2^*}{\sigma \sqrt{n}}\bigg)=\alpha\] where $\Phi$ denotes the cumulative distribution function of a Normal random variable. %Since \[\dfrac{n\theta_2^*-C_0}{\sigma \sqrt{n}}=-\dfrac{n\theta_2^*-C_0}{\sigma \sqrt{n}}\] and \[\dfrac{n\theta_1^*-C_0}{\sigma \sqrt{n}}=-\dfrac{n\theta_1^*-C_0}{\sigma \sqrt{n}}\], the two equations can be reduced to a single equation 
With the obtained critical constants, the $p$-value $P_f(\pmb{X})$ and its CDF can be calculated using Equations \eqref{normal_pval} and \eqref{normal_pval_cdf}, respectively.
%\[P(\pmb{X})=\Phi\bigg(\dfrac{n\theta_2^*-C_0}{\sigma \sqrt{n}}\bigg)-\Phi\bigg(\dfrac{C_0-n\theta_1^*}{\sigma \sqrt{n}}\bigg)\] 
%while the cumulative distribution function for this $p$-value $P(\pmb{X})$ is given by
%\[\mathbb{P}_{\theta}(P(\pmb{X})\leq t)=
%\Phi\bigg(\dfrac{n\theta_2^*-C_0}{\sigma \sqrt{n}}\bigg)-\Phi\bigg(\dfrac{C_0-n\theta_1^*}{\sigma \sqrt{n}}\bigg).\] 
\end{example}

%%%%%%%%%%%%%%%%%%%%%%%%%%%%%%%%%%%
\subsection{Effect of Noise on the Conservativity and Power Function}
\label{effectofnoise}
\begin{figure}[ht!]
\begin{center}
\includegraphics[width=16cm,height=8cm]{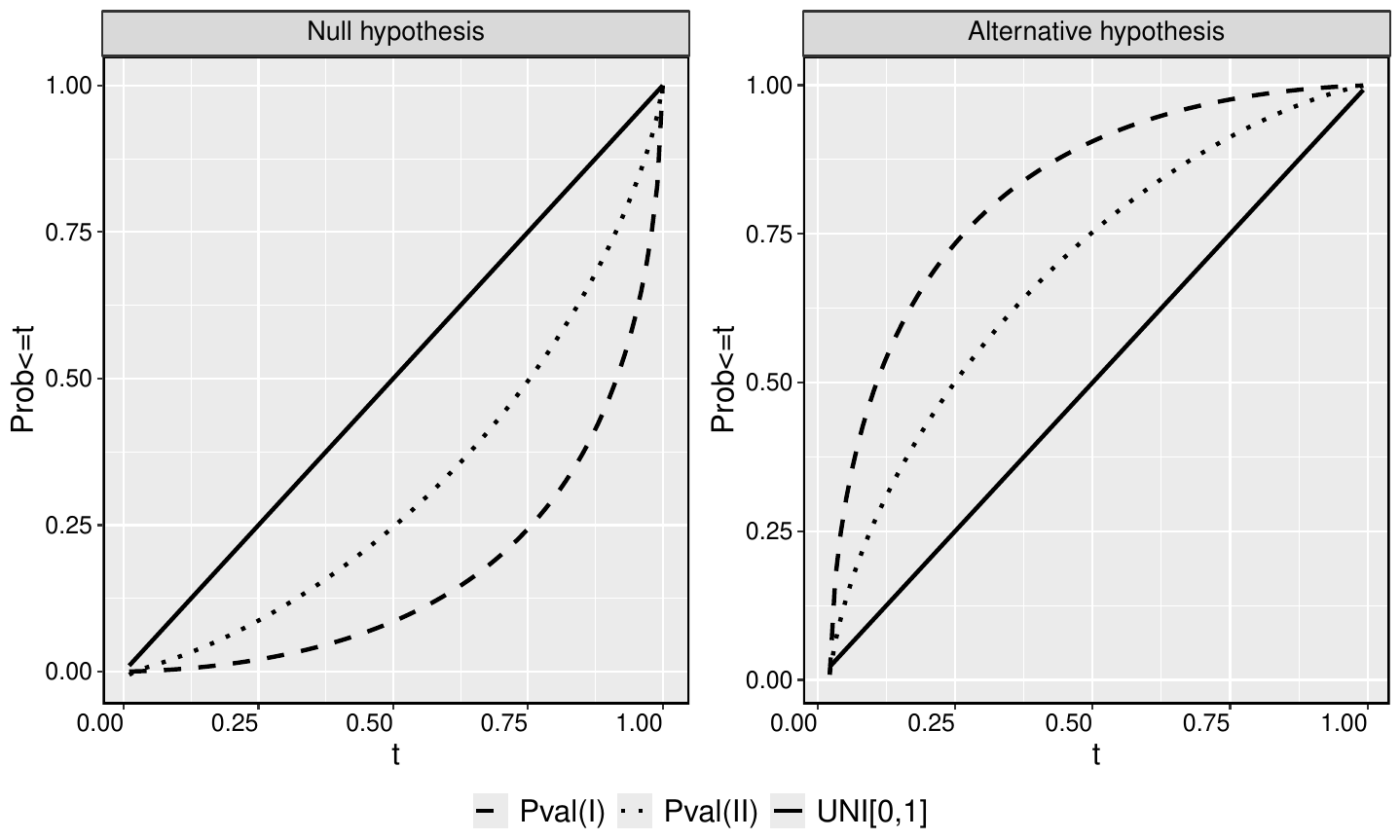}
\caption{An illustration of the effect of noise on the conservativity and power of the test based on the $p$-value using a sample of size $n=30$, equivalence margin $(\theta_1^*,\theta_2^*)=(1,4)$, and with different significance levels $t\in (0,1)$. We set the standard deviation $\sigma=2$ in $p$-val (I) and $\sigma=4$ in $p$-val (II). The true parameter under the null hypothesis is $0.5$, while the one under the alternative hypothesis is $1.5$.}
\label{Figure_05_equivalence_normal}
\end{center}
\end{figure}

%%%%%%%%%%%%%%%%%%%%%%%%%%%%%%%
In this section, we illustrate the effect of noise on the conservativity and power for the test based on the $p$-value. We plot Figure \ref{Figure_05_equivalence_normal} to illustrate this effect using a sample of size $n=30$, equivalence margin $(\theta_1^*,\theta_2^*)=(1,4)$, and with different significance levels $t\in (0,1)$. We set the standard deviation $\sigma=2$ in $p$-val (I) and $\sigma=4$ in $p$-val (II). The true parameter under the null hypothesis is $0.5$, while the one under the alternative hypothesis is $1.5$.

%check the power of the $p$-value in comparison to the one for the posterior probability if sd is larger than $\Delta$ or viceversa.

In Figure \ref{Figure_05_equivalence_normal}, using a larger standard deviation, the CDF for the $p$-value moves closer to the $\mathrm{Uni}(0,1)$ line under both the null hypothesis and the alternative hypothesis. This behavior implies that not only does the test based on the $p$-value become less conservative, but also loses power as the amount of noise in the data increases. These observations are also valid for randomized $p$-values such as those considered in \cite{hoang2021usage}, \cite{ochieng2024multiple}, and \cite{Ochieng2025thesis}. The only difference being a minor change under the null hypothesis. This decrease in power for the TOST procedure as variance increases is in line with the findings of \cite{berger1996bioequivalence}. 
The effect of noise here also illustrates the problem with nuisance parameters when calculating the $p$-value with unknown standard deviations. \cite{berger1994p} proposes some possible solutions for dealing with nuisance parameters. Alternatively, choosing the true parameter $\theta$ and the equivalence limit $(\theta_1^*,\theta_2^*)$ as a function of the standard deviation can also solve the problem.

%%%%%%%%%%%%%%%%%%%%%%%%%%%%%%%%%%%%%%%%
\subsection{Correlation Coefficient}
\label{effectofcorrelation}
%%%%%%%%%%%%%%%%%%%%%%%%%%%%%%%%%%%%%%%
In this section, we derive the correlation coefficient between the $p$-value and the posterior probability, as was done in \cite{dasgupta2000correlation} for one-sided tests. Our interest is to investigate whether the size of the equivalence margin affects this correlation coefficient. 
\begin{example}
\label{example:03}
As a continuation from Example \ref{example_normal}, recall that $X_1\sim N(\theta,\sigma^2)$. For the upper-tailed test in Equation \eqref{upper_tailed}, let $N(\theta_1^*,\tau^2)$ be the prior distribution $\pi(\theta)$ for the unknown parameter $\theta$. The posterior distribution $\pi(\theta|\bar{X})$ is then normally distributed with mean $(\theta_1^* \sigma^2+ n\tau^2 \bar{X})/(\sigma^2+n\tau^2)$ and variance $\tau^2 \sigma^2/(\sigma^2+n\tau^2)$. The posterior probability in this case is given by
\[
P_{r}(\theta\leq \theta_1^*|\bar{X},n)=1-\Phi\bigg(\dfrac{n\tau(\bar{X}-\theta_1^*)}{\sigma\sqrt{\sigma^2+n\tau^2}}\bigg),\]and the $p$-value is 
\[P_r(\pmb{X})=1-\Phi\bigg(\dfrac{\sqrt{n}(\bar{X}-\theta_1^*)}{\sigma}\bigg).
\]
Similarly, for the lower-tailed test in Equation \eqref{lower_tailed}, let $N(\theta_2^*,\tau^2)$ be the prior distribution $\pi(\theta)$ for the unknown parameter $\theta$. The posterior distribution $\pi(\theta|\bar{X})$ is then normally distributed with mean $(\theta_2^* \sigma^2+ n\tau^2 \bar{X})/(\sigma^2+n\tau^2)$ and variance $\tau^2 \sigma^2/(\sigma^2+n\tau^2)$. 
The posterior probability in this case is given by
\[P_{l}(\theta\geq \theta_2^*|\bar{X},n)=\Phi\bigg(\dfrac{n\tau(\bar{X}-\theta_2^*)}{\sigma\sqrt{\sigma^2+n\tau^2}}\bigg),\] and the $p$-value is 
\[P_l(\pmb{X})=\Phi\bigg(\dfrac{\sqrt{n}(\bar{X}-\theta_2^*)}{\sigma}\bigg).\] 
Note that we can also obtain these $p$-values from the posterior probabilities by holding the sample size $n$ constant and using a flatter prior by letting $\tau^2 \to \infty.$
With these two posterior probabilities, the overall posterior probability for the equivalence test is then 
\begin{equation}
P_b(\pmb{X})
%&=P_{r}(\theta\leq \theta_1^*|\bar{X},n)+P_{l}(\theta\geq \theta_2^*|\bar{X},n),\nonumber\\
%\nonumber\\
=1-\Phi\bigg(\dfrac{n\tau(\bar{X}-\theta_1^*)}{\sigma\sqrt{\sigma^2+n\tau^2}}\bigg)+\Phi\bigg(\dfrac{n\tau(\bar{X}-\theta_2^*)}{\sigma\sqrt{\sigma^2+n\tau^2}}\bigg).
\label{eq:pbayes_normalfinal}
\end{equation}
The overall $p$-value for the equivalence problem, on the other hand, is given by (see the details in Appendix \ref{equivalence_normal_p_val})
\begin{equation}
P_f(\pmb{X})=1-\bigg[\Phi\bigg(\dfrac{\sqrt{n}(\bar{X}-\theta_1^*)}{\sigma}\bigg)+\Phi\bigg(\dfrac{\sqrt{n}(\bar{X}-\theta_2^*)}{\sigma}\bigg)\bigg].
\label{eq:frequentist_normal_pval}
\end{equation}
%showing clearly, as is known, that these two measures of evidence are incompatible when dealing with an interval composite null of this type. 
\end{example}

The above example leads us to the following proposition concerning the correlation coefficient between these two measures of evidence.

\begin{proposition}
\label{prop:correlation}
The correlation coefficient between the posterior probability $P_b(\pmb{X})$ in Equation \eqref{eq:pbayes_normalfinal} and the $p$-value $P_f(\pmb{X})$ in Equation \eqref{eq:frequentist_normal_pval} when the hypothesis in Equation \eqref{hypothesis} is of interest is $\rho\{P_b(\pmb{X}),P_f(\pmb{X})\}=0$.
\end{proposition}
The proof of this proposition is given in Appendix \ref{prop:corr_equivalence_normal}. 
Finally, it is known \citep{arboretti2021unified} that the correlation coefficient between the partial tests in Equations \eqref{upper_tailed} and \eqref{lower_tailed} ranges from $-1$ when the size of the equivalence margin $\epsilon=0$ to approximately zero for sufficiently large margins. We state this in the following proposition.

\begin{proposition}
\label{prop:correlation_2}
The correlation coefficient between the partial tests in Equations \eqref{upper_tailed} 
 and \eqref{lower_tailed} is such that $\rho\in [-1,0)$ with the exact value depending on the size of the equivalence margin $\epsilon$.
\end{proposition}
The proof of this proposition for normal random variables when $\epsilon=0$ is provided in Appendix \ref{prop:correlation_2_proof}. We now extend our comparison of the properties of the $p$-value and the posterior probability to the case of multiple testing.

%%%%%%%%%%%%%%%%%%%%%%%%%%%%%%%%%%%%%%%%%%%%%
\section{Application to Multiple Testing}
\label{simulation}
%%%%%%%%%%%%%%%%%%%%%%%%%%%%%%%%%%%%%%%%%%%%%

\subsection{Introduction}

In the single hypothesis test case considered previously, the goal was to maximize power while controlling the Type I error rate. We extend this to the case of multiple tests, where the goal remains to maximize power for each test while controlling some compound error rate at the desired level. 
Recall that to test several hypotheses $k>1$, we express them as $H_j: \theta_j\notin \Delta$ versus $K_j:\theta_j\in \Delta$ where $\Delta$ is the range of values in the interval between $\theta_1^*$ and $\theta_2^*$, which shall remain constant for $j\in \{1,\ldots,k\}$ and $k$ is the multiplicity of the problem.
Therefore, of the $k$ hypotheses of interest, let $k_0$ denote the number of true null hypotheses and $k_1=k-k_0$ the number of false null hypotheses. Furthermore, assume that we have $R$ of the $k$ hypotheses rejected and $W$ as non-rejected null hypotheses. Also, denote the correct decisions by $S (U)$, which is the number of correct rejections (non-rejections), and the incorrect decisions by $V (T)$, which is the number of rejected null hypotheses (non-rejected alternative hypotheses). These two quantities $R$ and $W$ are observable random variables, while all the other quantities $U$, $V$, $T$, and $S$ are unknowns. We summarize this in Table \ref{tab:decision_table}. 
\begin{table}[ht!]
\centering  
\scalebox{.9}{
\begin{tabular}{ccc ccc c}
\hline
\hline
\rowcolor{gray}
%\cmidrule(lr){4-5} \cmidrule(lr){6-7}
%\cmidrule(lr){8-9}
\rowcolor{lightgray}  &$H$ not rejected & $H$ rejected    & Total \\
\midrule
$H$ true &$U$& $V$ & $k_0$\\
$H$ false &$T$& $S$ & $k-k_0$\\
Total &$W$& $R$ & $k$\\
\hline
\hline
\end{tabular}}
\caption{Summary classification of $k$ hypotheses tests.}
\label{tab:decision_table}
\end{table}

One popular error rate control when dealing with a large number of hypotheses in multiple testing is the false discovery rate (FDR) control \citep{benjamini1995controlling}. The goal of the FDR procedure is to minimize $Q=V/(R\vee 1)$, which is the proportion of false rejections, also called the false discovery proportion (FDP), and $a\vee b$ denotes $\max(a,b)$. Taking the maximum ensures that we do not have an expression of the kind $0/0$. Note that $Q=0$ if there is no rejection, that is, if $R=0$. Finding the expected value of $Q$ when $R=0$ becomes problematic. \cite{benjamini1995controlling} therefore proposed FDR control by finding the expectation $\mathbb{E}[V/R|R>0]*\mathbb{P}(R>0)$. Assuming that we have $k$ independent $p$-values from the $k$ hypotheses and that each follows a uniform $[0,1]$ distribution under the null hypothesis, the proposed procedure proceeds as follows. Order the $k$ $p$-values such that $p_{(1)}\leq p_{(2)}\leq\ldots\leq p_{(k)}$ and then reject the $D$ of the smallest $p$-values where 
\[D=\mathrm{max}\bigg\{j: p_{(j)}\leq \dfrac{j \alpha}{k}\bigg\}.\] 
The value $D$ is known as the ``deciding point" in the previous literature. This procedure controls the FDR at the level $(\alpha k_0/k)\leq\alpha$. This level of control remains the same for any number of true null hypotheses or any distribution of the $p$-values under the alternative hypothesis.
Replacing the value of $k$ by $k_0$ in the formula for $D$ can lead to an improvement in the power of this procedure. Since the value of $k_0$ is unknown in practice, we can use its estimate and the modified procedure, which has been called adaptive \citep{benjamini2000adaptive}, rejects $D$ of the smallest ordered $p$-values where now
\[D=\mathrm{max}\bigg\{j: p_{(j)}\leq \dfrac{j \alpha}{\widehat{k}_0}\bigg\}.\]

The power of the FDR procedure, which is the expected proportion of false null hypotheses out of the correct rejections, is given by $(R-V)/(k_1\vee 1)$. FDR methods are preferred when one is willing to tolerate a small proportion of erroneously rejected null hypotheses out of the rejected null hypotheses. We also note that FDR control implies familywise error rate (FWER) control when all $k$ null hypotheses are true. Furthermore, FWER control implies FDR control when $k_0<k$.

To carry out these simulation studies, we perform the TOST procedure by conducting the left- and right-tailed tests and then combining the two results. We proceed as follows. 
\begin{enumerate}
\item[(i)] For the right-tailed test, we generate the true alternative means $k_1$ from the distribution $N(\theta_1^*+\epsilon^*, \sigma^2)$ where $\epsilon^*$ is a small number. The true null means we generate from the distribution $N(\theta_1^*, \sigma^2)$. \item[(ii)] Similarly, for the left-tailed test, we generate the true alternative means $k_1$ from the distribution $N(\theta_2^*-\epsilon^*, \sigma^2)$  and the true null means we generate from the distribution $N(\theta_2^*, \sigma^2)$. 
\item[(iii)] We calculate the $Z$ statistic from the generated data and use this to calculate the $p$-values or the posterior probabilities. The combination of these two $p$-values (Equation \eqref{eq:frequentist_normal_pval}) or posterior probabilities (Equation \eqref{eq:pbayes_normalfinal}) is then used to calculate the power of the FDR procedure.  
\end{enumerate}

We set $\sigma^2=1$ throughout and use $k={1,000}$. We take the number of false null hypotheses $k_1$ as a sequence from $10$ to $990$ in an interval of $50$, and the level of significance is $\alpha=0.05$. Furthermore, in each simulation scenario, we specify the other values such as the sample size $n$, the equivalence margin $(\theta_1^*,\theta_2^*)$, and the prior standard deviation $\tau$. 
To simulate the average power of the FDR procedure, we repeat the three steps above $r={1,}000$ times and take the average. 

We compare the power of the FDR procedure when using the $p$-value and the posterior probability for these three cases.
\begin{enumerate}
\item[(i)] Different equivalence margins, where we investigate the behavior of the power of the FDR procedure for both the $p$-value and the posterior probability with a change in the equivalence margin. In practical situations, the equivalence margin is always chosen in advance and remains fixed for the entire experiment. The adjustments here are only for illustration.
\item[(ii)] Different values for the standard deviation $(\tau)$ of the prior distribution for the posterior probability, where we suspect that high values lead to increased power.
\item[(iii)] Different sample sizes in which we also suspect that the sample size affects the power of both measures of evidence much more when the equivalence margins are small compared to when they are large. 
\end{enumerate}

In what follows, we provide a detailed investigation of the three cases. The cases $(i)$ and $(iii)$ were also considered in \cite{qiu2010evaluation}.

%%%%%%%%%%%%%%%%%%%%%%%%%%%%%%%%%%%%
\subsection{Effect of the Equivalence Margins}
\begin{figure}[ht!]
\begin{center}
\includegraphics[width=16cm,height=14cm]{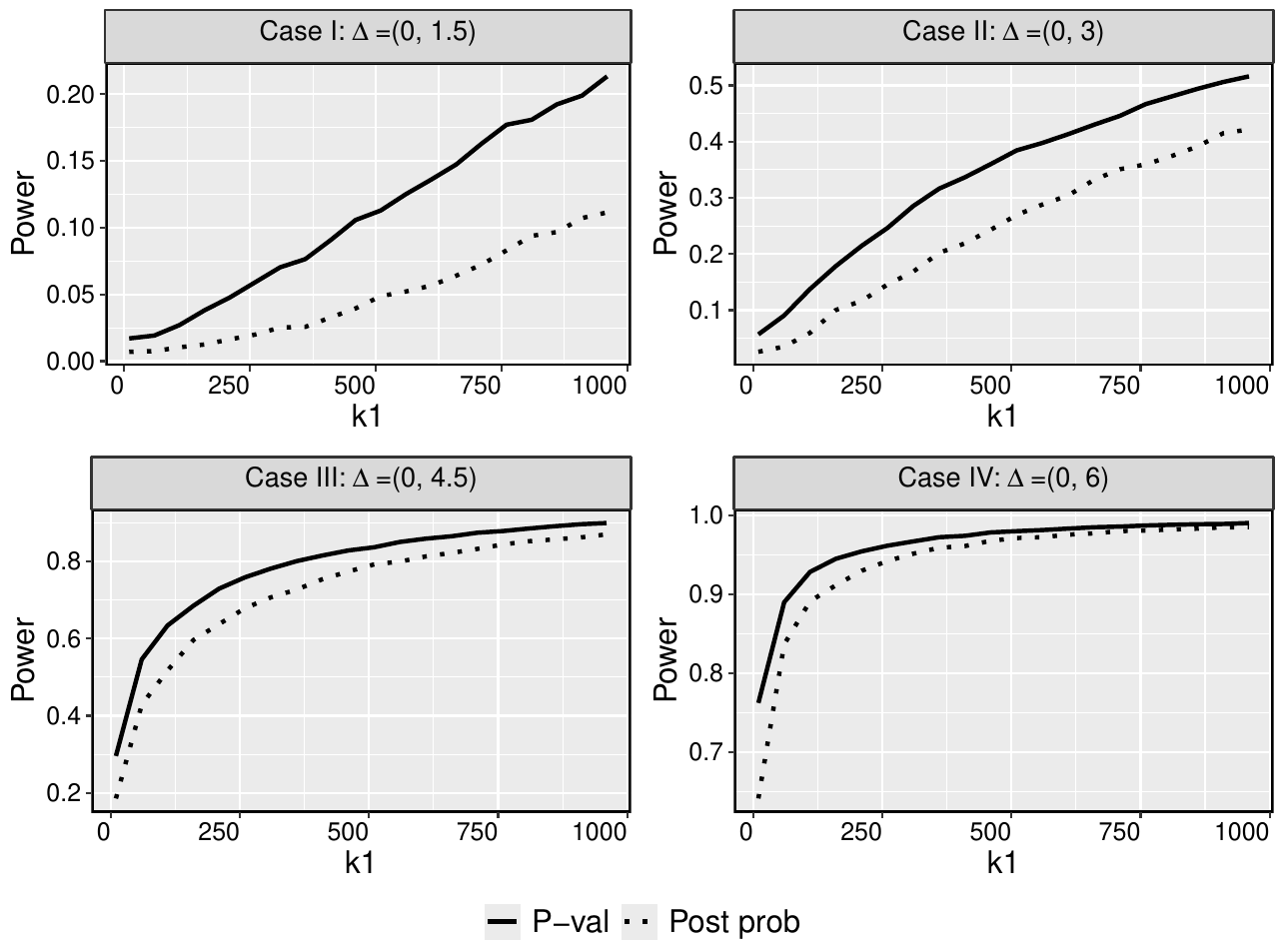}
\caption{An illustration of the effect of different equivalence margins on the power of the FDR procedure for the posterior probability in comparison to the $p$-value. We set the number of hypotheses at $k=1{,}000$, the level of significance at $t=0.05$, and the sample size at $n=100$. Furthermore, we set the number of replications at $r={1,}000$, the prior standard deviation at $\tau=0.25$, and use four different equivalence margins $(\theta_1^*,\theta_2^*)=(0,1.5), (0,3), (0,4.5)$, and $(0,6)$ for cases I to IV, respectively.}
\label{Figure_7_FDR_Power_versus_equivalence}
\end{center}
\end{figure}

%%%%%%%%%%%%%%%%%%%%%%%%%%%%%%%%%%%%
We plot Figure \ref{Figure_7_FDR_Power_versus_equivalence} to illustrate the effect of different equivalence margins on the FDR power for procedures based on the $p$-value and the posterior probability.  We set the sample size at $n=100$, the prior standard deviation at $\tau=0.25$, and use four different equivalence margins $(\theta_1^*, \theta_2^*)=(0,1.5), (0,3), (0,4.5)$, and $(0,6)$ for cases I to IV, respectively. 
In Figure \ref{Figure_7_FDR_Power_versus_equivalence}, the power of the FDR procedure for both measures of evidence increases as the equivalence margins increase. The gap between the two power curves also reduces as the equivalence margin increases, especially for a large number of false null hypotheses $k_1$. The FDR power when using the $p$-value is always higher than when using the posterior probability for any equivalence margin.

%%%%%%%%%%%%%%%%%%%%%%%%%%%%%%%%%%%%%%%%%%%%
\subsection{Effect of the Standard Deviation of the Prior Distribution}
\begin{figure}[ht!]
\begin{center}
\includegraphics[width=16cm,height=14cm]{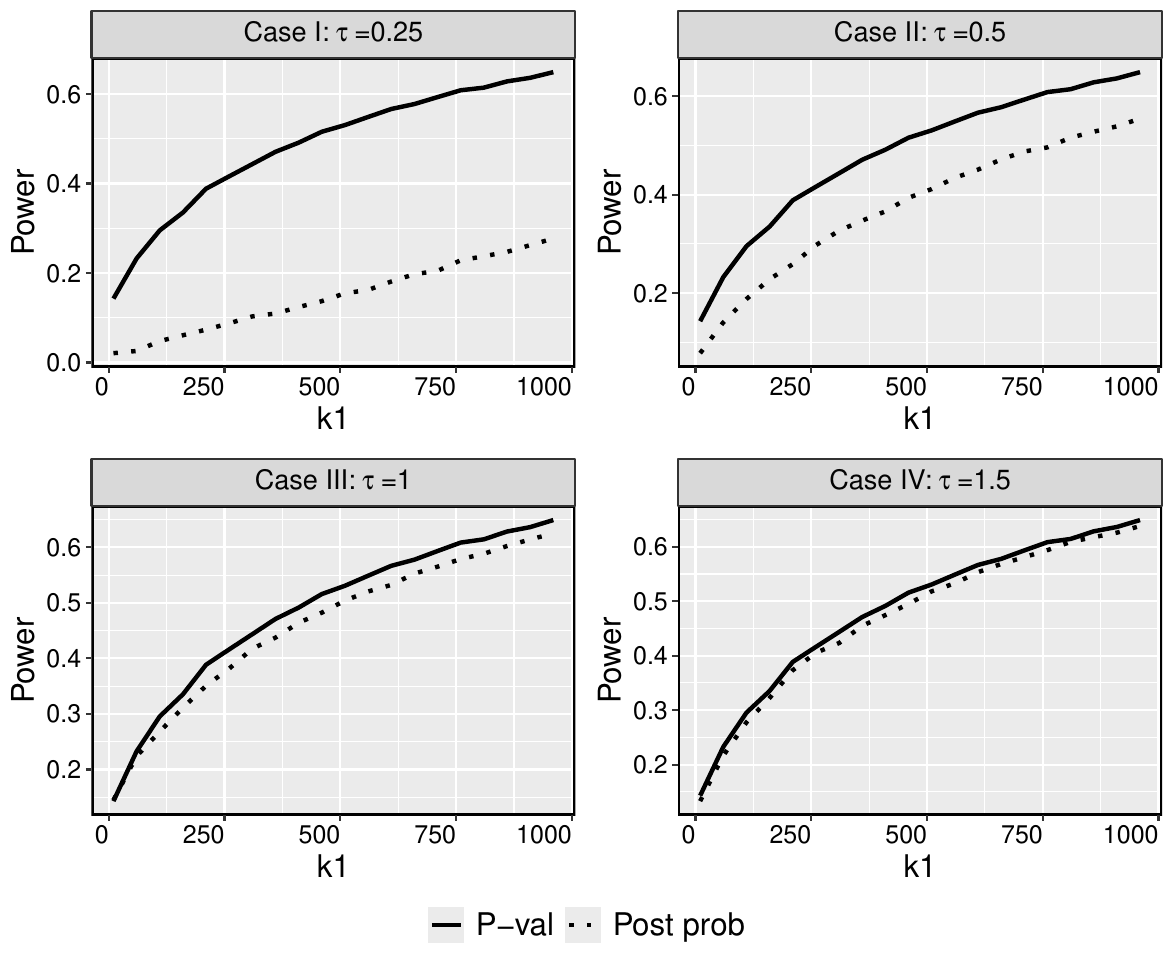}
\caption{An illustration of the effect of prior standard deviation on the power of the FDR for the posterior probability in comparison to the $p$-value. We set the number of hypotheses at $k=1{,}000$, the level of significance at $t=0.05$, and the sample size at $n=20$. Furthermore, we set the number of replications at $r={1,}000$, the equivalence margin at $(\theta_1^*,\theta_2^*)=(0,2)$, and use four different prior standard deviations $\tau\in\{0.25,0.5,1,1.5\}$. }
\label{Figure_06_fdr_power_prior_variance}
\end{center}
\end{figure}

%%%%%%%%%%%%%%%%%%%%%%%%%%%%%%%%%%%%%%%%%%%%
Here we repeat the above simulation studies using different values for the standard deviation of the prior distribution while keeping the sample size and equivalence margin constant. To illustrate the effect of this choice, we plot Figure \ref{Figure_06_fdr_power_prior_variance}. We set the sample size at $n=20$, the equivalence margin at $(\theta_1^*,\theta_2^*)=(0,2)$, and use four different values for the standard deviation of the prior distribution $\tau\in\{0.25,0.5,1,1.5\}$. 
In Figure \ref{Figure_06_fdr_power_prior_variance}, the power of the FDR procedure based on the $p$-value is higher than the one based on the posterior probability throughout.  
The FDR power based on the posterior probability moves closer to the one based on the $p$-value as the value $\tau$ increases. This trend remains the same as long as the value of $\tau$ is within the equivalence margin $(\theta_1^*,\theta_2^*)$. When $\tau\leq\theta_1^*$ or $\tau\geq\theta_2^*$, the power for the FDR procedure based on the posterior probability and the $p$-value coincides throughout.

%%%%%%%%%%%%%%%%%%%%%%%%%%%%%%%%%%%%%%%%%%%%%
\subsection{Effect of the Sample Size}
\begin{figure}[ht!]
\begin{center}
\includegraphics[width=16cm,height=14cm]{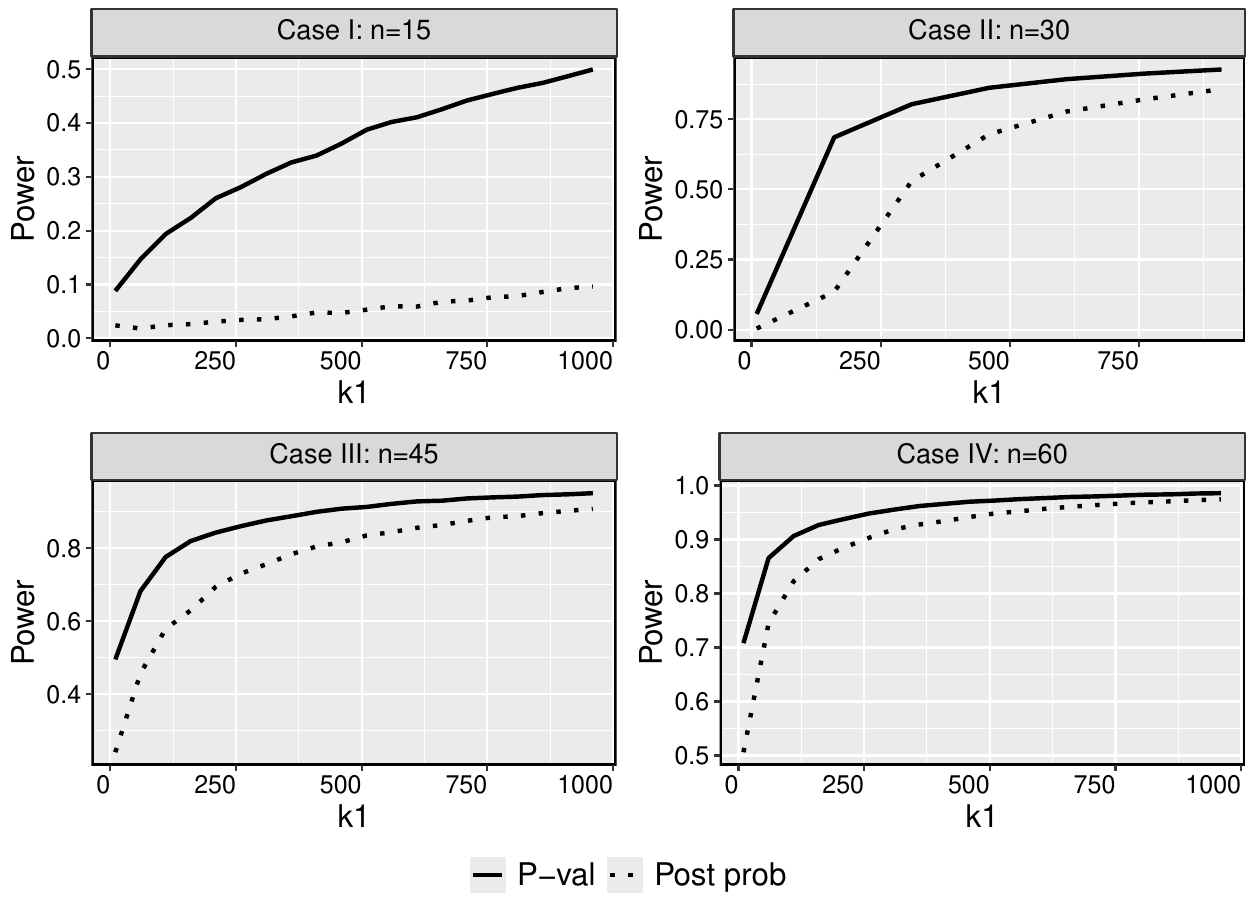}
\caption{An illustration of the effect of different sample sizes on the power of the FDR for the posterior probability in comparison to the $p$-value. We set the number of hypotheses at $k=1{,}000$, the level of significance at $t=0.05$, and the number of replications at $r={1,}000$. Furthermore, we set the equivalence margin at $(\theta_1^*,\theta_2^*)=(0,2)$, the prior standard deviation at $\tau=0.25$, and use four different sample sizes $n\in\{15,30,45,60\}$. }
\label{Figure_08_fdr_power_sample_size}
\end{center}
\end{figure}

%%%%%%%%%%%%%%%%%%%%%%%%%%%%%%%%%%%%%%%%%%%%%
In this case, we repeat the above simulation studies using different sample sizes while keeping both the prior standard deviation and the equivalence margin constant. To illustrate the effect of different sample sizes, we plot Figure \ref{Figure_08_fdr_power_sample_size}. We maintain the equivalence margin at $(\theta_1^*,\theta_2^*)=(0,2)$, the prior standard deviation at $\tau=0.25$, and use four different sample sizes $n\in\{15,30,45,60\}$. 
In Figure \ref{Figure_08_fdr_power_sample_size}, for small sample sizes (case I), the posterior probability has an extremely low power compared to the $p$-value. The power of both measures of evidence increases as the sample size increases. The power curve for the posterior probability still moves closer to the one for the $p$-value as the sample size increases. 
In this figure, both power curves increase monotonically with an increase in the number of false null hypotheses ($k_1$). This observation changes for small sample sizes with small equivalence margins or moderate sample sizes with large equivalence margins. We illustrate this in Case I of Figure \ref{Figure_9_FDR_Power_versus_small_equivalence} in the Appendix \ref{appendixB2}. 
At these small equivalence margins, we require large sample sizes to have reasonable power and have a monotonically increasing power function with increasing $k_1$. This claim is illustrated in Case II of Figure \ref{Figure_9_FDR_Power_versus_small_equivalence}. However, the power for the FDR based on the posterior probability is still very low compared to the $p$-value. Using a constant sample size while increasing the size of the equivalence margin also has the same problem (Case III in Figure \ref{Figure_9_FDR_Power_versus_small_equivalence}). 
Increasing both the sample size and the size of the equivalence margin leads to an increase in power and monotonic power functions for both measures of evidence (Case IV in Figure \ref{Figure_9_FDR_Power_versus_small_equivalence}). Generally, the posterior probability requires larger sample sizes to have reasonable power compared to the $p$-value.

%%%%%%%%%%%%%%%%%%%%%%%%%%%%%%%%%%
\section{Discussion}
\label{discussion}
%%%%%%%%%%%%%%%%%%%%%%%%%%%%%%%%%%

In this research, we have considered posterior probabilities and $p$-values to compare multiple parameters to a common equivalence margin. We have illustrated that in the binomial model with the conjugate beta prior distribution, the posterior probability is always less conservative and more powerful than the $p$-value for prior parameters less than one. 
As the prior parameters increase, the conservativity of the posterior probability increases, and for some values, the posterior probability is now slightly more conservative than the $p$-value. Similarly, the power of the posterior probability also reduces with an increase in the prior parameter, so that for some large values, the posterior probability is now less powerful than the $p$-value.
This observation implies that neither of the two tests is uniformly most powerful for all values of $\alpha$. Based on these results for a binomial model when using the posterior probability as a measure of evidence, it is advisable to use small parameter values in the beta prior distribution. 

Still, for the binomial model, we have also illustrated that the true parameter value that maximizes the power of the test based on the $p$-value is $\theta_{\mathrm{max}}=0.5$, which is the midpoint of $(0,1)$, the possible range for $\theta$. For the posterior probability, this parameter $\theta_{max}$ is less than, but close to $0.5$ and is exactly $0.5$ when using large values as prior parameters, which unfortunately leads to a drop in power. 
This trend persists regardless of the size of the equivalence margin. Furthermore, the trend remains the same for larger sample sizes, with the only difference being a slight decrease in power for large values of prior parameters. However, if the equivalence margin does not contain $0.5$ and we have $ \theta_2^*<0.5$, then $\theta_{max}$ appears close to the midpoint of the equivalence margin and only moves closer to $\theta_2^*$ as the values of the prior parameters increase.
Also, if the equivalence margin does not contain $0.5$ and we have $\theta_1^*>0.5$, then $\theta_{max}$ still appears close to the midpoint of the equivalence margin, but now moves closer to $\theta_1^*$ as the values of the prior parameters increase.  

Again, for the binomial model, we have illustrated that holding one prior parameter constant while increasing the other one leads to a change in the maximum parameter $\theta_{max}$ and the power curve for the posterior probability. This shift is to the left or right of the power curve for the $p$-value depending on which parameter is kept constant. This shift also led to power loss, but it was not as severe as the one in the previous paragraph. 
The parameter value that maximizes the power function for the posterior probability can only be equal to the one for the $p$-value without a power loss if the prior distribution is non-symmetric.
Generally, for small prior parameters as illustrated in Figures \ref{Figure_max_power}, \ref{Figure_changing_prior_b}, and \ref{Figure_max_power_alpha}, the test based on the posterior probability always has more power than the one based on the $p$-value when the true parameter under the alternative hypothesis is lower than $\theta_{\mathrm{max}}=0.5$. 
%\Cref{Figure_max_power}

We also conducted additional simulation studies to compare the type I error rate and the power of the posterior probability and the $p$-value in discrete models. We have done this for varying equivalence margins and sample sizes. We still found that the posterior probability with small prior parameters outperforms the $p$-value for any sample size and any equivalence margin. The difference between the two measures of evidence is minimal for wide equivalence margins, but becomes more apparent for small margins. For large prior parameter values, there is minimal difference between the two procedures. We relate our results to those of \cite{shi2021reconnecting}, who considered one-sided tests with a noninformative prior distribution. They found that the posterior probability and the $p$-value are asymptotically equivalent. For the two-sided test procedure, the $p$-value equals a modified posterior probability. 

We have also investigated the effect of the standard deviation on the conservativity and power function of the test based on the $p$-value for the normal model. As the standard deviation increases, the $p$-value becomes less conservative and also less powerful. As in \cite{munk1999note}, it would be interesting to investigate for which standard deviation values the test based on the $p$-value has more power than the one based on the posterior probability and vice versa, thus demonstrating that none is a uniformly most powerful (UMP) test. Furthermore, we know that the TOST procedure is biased, especially as the standard deviation increases \citep{munk2000unbiased}. It would be interesting to compare the bias of this procedure when using the $p$-value and posterior probabilities. 

Ideally, the power of the TOST procedure should approach zero uniformly in $\theta$ as the standard deviation increases \citep{brown1997unbiased,munk2000unbiased}. We suspect that our $p$-value is unbiased, and that is why the power function tends to $\alpha$ as the standard deviation increases, as illustrated in Figure \ref{Figure_05_equivalence_normal}. Distributions also exist that have a noncentrality parameter \citep{muller1990power}. It would be interesting to investigate the effect of this noncentrality parameter on the conservativity and power of the test based on the $p$-values and posterior probabilities. 
%\newline

We also have similar findings to those in \cite{berger1987testing}, who considered an interval null hypothesis. They found that the posterior probability under the null hypothesis also tends to $\alpha$ as $n$ tends to infinity. Furthermore, at large sample sizes, the problem essentially reduces to a one-sided testing problem where the $p$-value and the posterior probability are known to coincide.

We also extend this comparison to multiple tests, where we compare the power of the FDR procedure when using the $p$-value and the posterior probability. We compare the power functions for different equivalence margins, sample sizes, and standard deviation of the prior distribution. The FDR power based on the $p$-value exceeds that based on the posterior probability throughout the simulation studies. The power function for both measures of evidence always increases monotonically with an increase in the number of false null hypotheses, unless the equivalence margin is too narrow and the sample size is too low. The power function also increases with an increase in the sample size or an increase in the size of the equivalence margin.

As mentioned in Section \ref{introduction}, there is a need to develop procedures that compare a single or multiple parameters among many groups. In this research, we have compared a single parameter among several groups. One could extend this work to consider the comparison of multiple parameters among multiple groups, for example, conducting marginal tests and then using some $p$-value combination method for the multiple tests. We prefer this avenue of conducting marginal tests since no other method is available to compare several parameters among multiple groups that is more powerful than the TOST procedure \citep{pallmann2017simultaneous}. However, this approach can lead to conservative tests unless a $p$-value is available that achieves exact uniformity under the null hypothesis.

%\begin{example}[Publication bias and the failure of classical inference]
%\label{exm:publication_bias}
%\end{example}

%\begin{definition}[Selective dominance]
%\label{def:selective_dominance}  
%\end{definition}

%\begin{theorem}[Selective dominance and error control]
%\label{thm:adjustment} 
%\end{theorem}

%\begin{corollary}[Testing the winner]
%\label{cor:cond}
%\end{corollary}

%\begin{proof}[Proof sketch]      
%\end{proof}

%\section*{Acknowledgements}
%The initial draft of this article was presented at a discussion meeting with my former colleagues during my PhD. I would therefore like to thank these former colleagues for the helpful discussions that led to a better presentation of this article.

\bibliographystyle{plainnat}
\bibliography{ref.bib}

\begin{appendix}

\section{Additional Comments and Proofs}

\subsection{Calculating the $p$-Value for the Equivalence Test}
\label{equivalence_normal_p_val}
We provide in this section details concerning the calculation of the equivalence $p$-value mentioned in Equation \eqref{eq:frequentist_normal_pval}. Taking ideas from \cite{berger1987testing}, we first rewrite the hypothesis in
Equation \eqref{hypothesis} as $H:|\theta-\theta_0|\geq\epsilon$ versus $K:|\theta-\theta_0|<\epsilon$, where $\epsilon$ is a small positive constant. 
Note that in the multiple test case, we always use the same constant $\epsilon$, and therefore we are dealing with the \textit{average equivalence} criterion. 
We define the overall $p$-value as
\[P_f(\pmb{X})=
P_{\theta}(|T|\leq |t|),\]
%\sup_{\theta:|\theta-\theta_0|\geq \epsilon} 
for $\theta:|\theta-\theta_0|\geq \epsilon$.
Assuming $\bar{X}\sim N(\theta, \sigma/\sqrt{n})$ where $\sigma^2$ is known and choosing $T(\bar{X})=\dfrac{\sqrt{n}(\bar{X}-\theta_0)}{\sigma}$ as our test statistic and $t=T(\bar{x})$, we have the $p$-value calculated as
\[P_f(\pmb{X})=P\bigg(\dfrac{\sqrt{n}(\bar{X}-\theta_0)}{\sigma}\leq |t|\bigg)-P\bigg(\dfrac{\sqrt{n}(\bar{X}-\theta_0)}{\sigma}\leq -|t|\bigg).\]
Making $\bar{X}$ the subject in $P_f(\pmb{X})$ and standardizing using the distribution of $\bar{X}$, we have
\[P_f(\pmb{X})=\Phi\bigg(|t|+\dfrac{\epsilon\sqrt{n}}{\sigma} \bigg)+\Phi\bigg( |t|-\dfrac{\epsilon\sqrt{n}}{\sigma}\bigg)-1.\]
Since $t$ is just the realized value of $T$, we can substitute $t=\dfrac{\sqrt{n}(\bar{x}-\theta_0)}{\sigma}$ into $P_f(\pmb{X})$ and simplify. Therefore, we have
\[P_f(\pmb{x})=\Phi\bigg(
\dfrac{\sqrt{n}(\bar{x}-\theta_1^*)}{\sigma}
\bigg)+\Phi\bigg( \dfrac{\sqrt{n}(\bar{x}-\theta_2^*)}{\sigma}
 \bigg)-1,\]
 where $\theta_1^*$ and $\theta_2^*$ denote $\theta_0-\epsilon$ and $\theta_0+\epsilon$, respectively. This equation is equivalent to the one in \eqref{eq:frequentist_normal_pval} when the true parameter $\theta$ under the null hypothesis is such that $\theta=\theta_1^*$ or $\theta=\theta_2^*$, which are the LFC parameters in this case.

\subsection{Proof of Proposition \ref{prop:correlation}}
\label{prop:corr_equivalence_normal}
%%%%%%%%%%%%%%%%%%%%%%%%%%%%%%%%%
From Example \ref{example:03}, let 
$t_i=\dfrac{n\tau(\bar{X}-\theta_i^*)}{\sigma\sqrt{\sigma^2+n\tau^2}}$ and $t_i^*=\dfrac{\sqrt{n}(\bar{X}-\theta_i^*)}{\sigma}$
for $i=1,2$.
The covariance between these two measures of evidence is given by 
\begin{eqnarray*}
\mathrm{Cov}\{P_b(\pmb{X}),P_f(\pmb{X})\}&=&\bigg\{\mathbf{E}\bigg[\Phi(t_1^*)\Phi(t_1)\bigg]-\mathbf{E}\bigg[\Phi(t_1^*)\bigg]\mathbf{E}\bigg[\Phi(t_1)\bigg]\bigg\}\\
&-&\bigg\{\mathbf{E}\bigg[\Phi(t_1^*)\Phi(t_2)\bigg]-\mathbf{E}\bigg[\Phi(t_1^*)\bigg]\mathbf{E}\bigg[\Phi(t_2)\bigg]\bigg\}\\
&+&
\bigg\{\mathbf{E}\bigg[\Phi(t_1)\Phi(t_2^*)\bigg]-\mathbf{E}\bigg[\Phi(t_1)\bigg]\mathbf{E}\bigg[\Phi(t_2^*)\bigg]\bigg\}\\
&-&
\bigg\{\mathbf{E}\bigg[\Phi(t_2^*)\Phi(t_2)\bigg]-\mathbf{E}\bigg[\Phi(t_2^*)\bigg]\mathbf{E}\bigg[\Phi(t_2)\bigg]\bigg\}.
\end{eqnarray*}

%Since the marginal distribution of $\bar{X}$ is normal with mean $\theta_i^*$ and variance $\tau^2+\sigma^2/n$ for $i=1,2$, 
We work out the above equation step by step as follows. We have for the first curly bracket
\[\mathbf{E}\bigg[\Phi(t_1^*)\Phi(t_1)\bigg]-\mathbf{E}\bigg[\Phi(t_1^*)\bigg]\mathbf{E}\bigg[\Phi(t_1)\bigg]\]
\[=\mathbb{E}\bigg[\Phi\bigg(\dfrac{\sqrt{\sigma^2+n\tau^2}}{\sigma}Z\bigg)\Phi\bigg(\dfrac{\sqrt{n}\tau }{\sigma}Z\bigg)\bigg]-\mathbb{E}\bigg[\Phi\bigg(\dfrac{\sqrt{\sigma^2+n\tau^2}}{\sigma}Z\bigg)\bigg]\mathbb{E}\bigg[\Phi\bigg(\dfrac{\sqrt{n}\tau }{\sigma}Z\bigg)\bigg],
\] where $Z\sim N(0,1).$
\[=\bigg(\dfrac{1}{4}\bigg)+\dfrac{1}{2\pi}\mathrm{arcsin}\bigg[\dfrac{\sqrt{n}\tau}{\sqrt{n\tau^2+2\sigma^2}}\bigg]-\bigg(\dfrac{1}{2}\bigg)*\bigg(\dfrac{1}{2}\bigg),\]
\[=\dfrac{1}{2\pi}\mathrm{arcsin}\bigg[\dfrac{\sqrt{n}\tau}{\sqrt{n\tau^2+2\sigma^2}}\bigg].\]
The other equations inside the curly brackets will give a similar result and so $\mathrm{Cov}\{P_b(\pmb{X}),P_f(\pmb{X})\}=0$. With this, it follows that the correlation coefficient between these two measures of evidence is zero, which is what we needed to prove.

%%%%%%%%%%%%%%%%%%%%%%%%%%%%%%%%%
\subsection{Proof of Proposition \ref{prop:correlation_2}}
\label{prop:correlation_2_proof}
%%%%%%%%%%%%%%%%%%%%%%%%%%%%%%%%%
%\begin{proof}
To prove this proposition, we need to calculate the correlation coefficient between the two $p$-values 
\[P_r(\pmb{X})=1-\Phi\bigg(\dfrac{\sqrt{n}(\bar{X}-(\theta_0-\epsilon))}{\sigma}\bigg)
\ \mathrm{and}\ P_l(\pmb{X})=\Phi\bigg(\dfrac{\sqrt{n}(\bar{X}-(\theta_0+\epsilon))}{\sigma}\bigg).\]
For this, we again follow the proof in \cite{dasgupta2000correlation}. We thus have the following.
\begin{align*}
\mathrm{Cov}\{P_r(\pmb{X}),P_l(\pmb{X})\}&=\mathbb{E}\bigg\{\Phi\bigg(\dfrac{\sqrt{n}(\bar{X}-(\theta_0-\epsilon))}{\sigma}\bigg)\bigg\}\mathbb{E}\bigg\{\Phi\bigg(\dfrac{\sqrt{n}(\bar{X}-(\theta_0+\epsilon))}{\sigma}\bigg)\bigg\}\\
&-\mathbb{E}\bigg\{\Phi\bigg(\dfrac{\sqrt{n}(\bar{X}-(\theta_0-\epsilon))}{\sigma}\bigg)\Phi\bigg(\dfrac{\sqrt{n}(\bar{X}-(\theta_0+\epsilon))}{\sigma}\bigg)\bigg\},\\
&=\mathbb{E}\bigg\{\Phi\bigg(Z+\dfrac{\sqrt{n}\epsilon}{\sigma}\bigg)\bigg\}\mathbb{E}\bigg\{\Phi\bigg(Z-\dfrac{\sqrt{n}\epsilon}{\sigma}\bigg)\bigg\}\\
&-\mathbb{E}\bigg\{\Phi\bigg(Z+\dfrac{\sqrt{n}\epsilon}{\sigma}\bigg) \Phi\bigg(Z-\dfrac{\sqrt{n}\epsilon}{\sigma}\bigg)\bigg\},\\
\intertext{where we have used the fact that $\bar{X}\sim N(\theta_0, \sigma^2/n)$ in the standardization process. 
We use $Z$ to denote a standard normal random variable throughout the proof. When $\epsilon=0$, we have}
\mathrm{Cov}\{P_r(\pmb{X}),P_l(\pmb{X})\}&=
\mathbb{E}\bigg\{\Phi(Z)\bigg\}\mathbb{E}\bigg\{\Phi(Z)\bigg\}-\mathbb{E}\bigg\{\Phi(Z) \Phi(Z)\bigg\},\\
&=-\bigg(\dfrac{1}{2\pi}\bigg)\mathrm{arcsin}\bigg(\dfrac{1}{2}\bigg).
\intertext{To find the variances, we have that}
\mathrm{Var}\{P_r(\pmb{X})\}&=
\mathbb{E}\bigg\{\Phi\bigg(Z-\dfrac{\sqrt{n}\epsilon}{\sigma}\bigg)\Phi\bigg(Z-\dfrac{\sqrt{n}\epsilon}{\sigma}\bigg)\bigg\}-\mathbb{E}\bigg\{\Phi\bigg(Z-\dfrac{\sqrt{n}\epsilon}{\sigma}\bigg)\bigg\}\mathbb{E}\bigg\{ \Phi\bigg(Z-\dfrac{\sqrt{n}\epsilon}{\sigma}\bigg)\bigg\}.
\intertext{Again when $\epsilon=0$, we have}
\mathrm{Var}\{P_r(\pmb{X})\}&=
\mathbb{E}\bigg\{\Phi(Z)\Phi(Z)\bigg\}-\mathbb{E}\bigg\{\Phi(Z)\bigg\}\mathbb{E}\bigg\{ \Phi(Z)\bigg\},\\
&=\bigg(\dfrac{1}{2\pi}\bigg)\mathrm{arcsin}\bigg(\dfrac{1}{2}\bigg).
\intertext{Similarly,}
\mathrm{Var}\{P_l(\pmb{X})\}&=\bigg(\dfrac{1}{2\pi}\bigg)\mathrm{arcsin}\bigg(\dfrac{1}{2}\bigg)\ \text{when $\epsilon=0$}.
\intertext{With the above covariance and variances, one can readily see that}
\rho\{P_r(\pmb{X}),P_l(\pmb{X})\}&=-1\ \text{when $\epsilon=0$, which is what we wanted to prove.}
\end{align*}
%\end{proof}

%%%%%%%%%%%%%%%%%%%%%%%%%%%%%%%%%%%%%
\subsection{Correlation Coefficient in the Two-Sided Problem}
%%%%%%%%%%%%%%%%%%%%%%%%%%%%%%%%%%%%%
For the sake of completeness, we also calculate the correlation coefficient between the posterior probability and the $p$-value in the two-sided problem given by 
\[H:\theta=\theta_0\ \mathrm{versus}\ K:\theta\neq \theta_0.\]
We still assume that $X_1\sim N(\theta,\sigma^2)$ and that $N(\theta_0,\tau^2)$ is the prior distribution $\pi(\theta)$ for the unknown parameter $\theta$. The posterior distribution $\pi(\theta|\bar{X})$ is then normally distributed with mean $(\theta_0 \sigma^2+ n\tau^2 \bar{X})/(\sigma^2+n\tau^2)$ and variance $\tau^2 \sigma^2/(\sigma^2+n\tau^2)$. 
In this case, the posterior probability is given by 
\[P_b(\pmb{X})=2\bigg[1-\Phi\bigg(\dfrac{n\tau(\bar{X}-\theta_0)}{\sigma\sqrt{\sigma^2+n\tau^2}}\bigg)\bigg],\]
and the $p$-value is
\[P_f(\pmb{X})=2\bigg[1-\Phi\bigg(\dfrac{\sqrt{n}(\bar{X}-\theta_0)}{\sigma}\bigg)\bigg].\]
%To derive the correlation coefficient, we again follow the calculations in \cite{dasgupta2000correlation}.
%Thus we have that the covariance is given by
%\begin{eqnarray*}
%\mathrm{Cov}\{P_b(\pmb{X}),P_f(\pmb{X})\}&=&4\mathbb{E}\bigg\{\Phi\bigg(\dfrac{n\tau(\bar{X}-\theta_0)}{\sigma\sqrt{\sigma^2+n\tau^2}}\bigg)\Phi\bigg(\dfrac{\sqrt{n}(\bar{X}-\theta_0)}{\sigma}\bigg)\bigg\}\\
%&-&4\mathbb{E}\bigg\{
%\Phi\bigg(\dfrac{n\tau(\bar{X}-\theta_0)}{\sigma\sqrt{\sigma^2+n\tau^2}}\bigg)
%\bigg\}\mathbb{E}\bigg\{\Phi\bigg(\dfrac{\sqrt{n}(\bar{X}-\theta_0)}{\sigma}\bigg)\bigg\}.
%\end{eqnarray*}
%Since the marginal distribution of $\bar{X}$ is the normal distribution with mean $\theta_0$ and variance $(\sigma^2/n)+\tau$. The above equation can be expressed as
%\begin{eqnarray*}
%\mathrm{Cov}\{P_b(\pmb{X}),P_f(\pmb{X})\}&=&4\mathbb{E}\bigg\{\Phi\bigg(\dfrac{\sqrt{\sigma^2+n\tau^2}}{\sigma}Z\bigg)\Phi\bigg(
%\dfrac{\sqrt{n}\tau}{\sigma}Z
%\bigg)\bigg\}\\
%&-&4\mathbb{E}\bigg\{
%\Phi\bigg(\dfrac{\sqrt{\sigma^2+n\tau^2}}{\sigma}Z
%\bigg)
%\bigg\}\mathbb{E}\bigg\{\Phi\bigg(
%\dfrac{\sqrt{n}\tau}{\sigma}Z
%\bigg)\bigg\},
%\end{eqnarray*}
%where $Z$ is a standard normal random variable.
%\[\therefore \mathrm{Cov}\{P_b(\pmb{X}),P_f(\pmb{X})\}=\bigg(\dfrac{2}{\pi}\bigg)\mathrm{arcsin}\bigg(\dfrac{\sqrt{n}\tau}{\sqrt{2\sigma^2+n\tau^2}}\bigg).\]
%Similarly
%\[\mathrm{Var}\{P_f(\pmb{X})\}=\bigg(\dfrac{2}{\pi}\bigg)\mathrm{arcsin}\bigg(\dfrac{\sigma^2+n\tau^2}{2\sigma^2+n\tau^2}\bigg),\] and
%\[\mathrm{Var}\{P_b(\pmb{X})\}=\bigg(\dfrac{2}{\pi}\bigg)\mathrm{arcsin}\bigg(\dfrac{n\tau^2}{\sigma^2+n\tau^2}\bigg).\]
We simplify this calculation by noting that for two random variables $X$ and $Y$, the correlation coefficient $\rho(aX, bY)=\rho(X, Y)$ where $a$ and $b$ are positive constants. Therefore, 
\[\rho\{P_b(\pmb{X}),P_f(\pmb{X})\}=\dfrac{\mathrm{arcsin}(\sqrt{w/(2-w)})}{\sqrt{\mathrm{arcsin}(w)\mathrm{arcsin}(1/(2-w))}},\]
where $w=n\tau^2/(\sigma^2+n\tau^2)$.
The correlation coefficient between the $p$-value and the posterior probability in the two-sided hypothesis test is equal to the one for the upper-tailed test calculated in \cite{dasgupta2000correlation}.

%%%%%%%%%%%%%%%%%%%%%%%%%%%%%%%%%%%%
\section{Additional Simulations and Figures}
\label{appendixB}
%%%%%%%%%%%%%%%%%%%%%%%%%%%%%%%%%%%%

%%%%%%%%%%%%%%%%%%%%%%%%%%%%%%%%%%
\subsection{Discrete Models}
%%%%%%%%%%%%%%%%%%%%%%%%%%%%%%%%%%
In this Section, we provide additional simulation results under different scenarios to compare the posterior probability to the $p$-value in discrete models. We now consider the simulations under two settings, which are (i) different sample sizes and (ii) different equivalence margins. We repeat the simulations $r=10{,}000$ times and the critical value is taken randomly in the interval $(0,n)$. We compare the two measures of evidence based on their type I error rate control and power.
We generate Table \ref{tab:sample_size} where we have used the equivalence margin $(\theta_1^*,\theta_2^*)=(0.25,0.75)$, prior parameters $(p,q)=(0.5,0.5), (1,1),$ and $(3,3)$, and varying sample sizes $n$ ranging from $20$ to $80$ with an interval of $10$. 
\begin{table}[ht!]
    \centering  
    \scalebox{.9}{
\begin{tabular}{ccc cc}
\hline
\hline
\rowcolor{gray}
 & &  \multicolumn{1}{c}{$\mathrm{Beta}(0.5,0.5)$} & 
  \multicolumn{1}{c}{$\mathrm{Beta}(1,1)$} & 
   \multicolumn{1}{c}{$\mathrm{Beta}(3,3)$}\\
%\cmidrule(lr){3-5}
\rowcolor{gray}
 $\textbf{n}$  & $\textbf{$P_f$}$   & \textbf{$P_b$}  & \textbf{$P_b$}  & \textbf{$P_b$}  \\
\midrule
\rowcolor{lightgray}
 20 & 0.0382  & 0.0693  & 0.0192  & 0.0734 \\
    &  0.3455 & 0.5257  &  0.2258 & 0.1443       \\
\rowcolor{lightgray}
 30 & 0.0197  & 0.0365  & 0.0281  & 0.0731 \\
    & 0.3972  & 0.5455  & 0.4580 &  0.2962       \\
 \rowcolor{lightgray}
40  & 0.0240  & 0.0403  & 0.0340 & 0.0693 \\
    &  0.5529 & 0.6752  & 0.6049 &  0.4270       \\
\rowcolor{lightgray}
50  & 0.0271  & 0.0421  & 0.0354  & 0.0654 \\
    &  0.6625 & 0.7570 & 0.6995  & 0.5469       \\
\rowcolor{lightgray}
60  & 0.0278  & 0.0422  & 0.0358 & 0.0610 \\
    &  0.7458 & 0.8160 & 0.7684 & 0.6450       \\
\rowcolor{lightgray}
70  & 0.0279  & 0.0417  & 0.0354  & 0.0579 \\
    &  0.8026 & 0.8611 & 0.8220  &  0.7232       \\
\rowcolor{lightgray}
80  & 0.0479  & 0.0638  & 0.0349  & 0.0540 \\
    &  0.8939 & 0.9340 & 0.8707 &  0.7797      \\
\hline
\hline
\end{tabular}}
 \caption{Type I error rate (upper row) and power (lower row) simulations for the $p$-value and the posterior probability using different sample sizes $n$, equivalence margin $(\theta_1^*,\theta_2^*)=(0.25,0.75)$, and prior parameters $(p,q)=(0.5,0.5),(1,1),$ and $(3,3)$. We let $\theta=0.4$ be the true parameter under the alternative hypothesis for the power simulation.}
    \label{tab:sample_size}
\end{table}
From Table \ref{tab:sample_size}, the posterior probability performs better than the $p$-value in controlling the type I error rate. This observation is valid for most sample sizes and especially for small prior parameters. For large values of the prior parameters, the performance of the posterior probability deteriorates compared to the $p$-value. 
The power of the test based on the posterior probability with low values for the prior parameters, such as $\mathrm{Beta}(0.5,0.5)$, exceeds that of the $p$-value for all sample sizes. The power also decreases with an increase in the value of the prior parameters but is still comparable to that of the $p$-value. 

These observations on the posterior probability remain the same even when using a nonsymmetric prior and only increasing one of the parameters while holding the other one constant. We make a similar observation when a nonsymmetric prior is used, and we increase both parameters simultaneously. In a related research, \cite{zaslavsky2013bayesian} also found that for an upper-tailed test, the posterior probability when using a flat prior ($p=q=1$) is always less conservative than the $p$-value. 

We are also interested in comparing the type I error rate and power when the sample size is constant while the equivalence margin is not. 
%Although the equivalence margin is always chosen in advance and remains fixed for the entire experiment as earlier explained, our adjustments here are just for illustrations and not to suggest that equivalence margins should be adjusted halfway through the experiment. 
For this, we present our results in Table \ref{tab:different_margins} where we have used a sample of size $n=35$, prior parameters $(p,q)=(0.5,0.5), (1,1)$, and $(3,3)$ with different equivalence margins $\Delta=\theta_2-\theta_1$.
\begin{table}[ht!]
\centering  
\scalebox{.9}{
\begin{tabular}{ccc ccc c}
\hline
\hline
\rowcolor{gray}
& & & & \multicolumn{1}{c}{$\mathrm{Beta}(0.5,0.5)$} & 
\multicolumn{1}{c}{$\mathrm{Beta}(1,1)$} & 
\multicolumn{1}{c}{$\mathrm{Beta}(3,3)$}\\
%\cmidrule(lr){4-5} \cmidrule(lr){6-7}
%\cmidrule(lr){8-9}
\rowcolor{gray} $\theta_1$ &$\theta_2$ & $\Delta$    & $P_f$   & $P_b$  & $P_b$ & $P_b$\\
\midrule
\rowcolor{lightgray} 
0.150&0.850&0.70&0.0278&0.0490&0.0346&0.0720\\
&&&0.9405&0.9751&0.9574&0.8683\\
\rowcolor{lightgray}
0.175 & 0.825 & 0.65  & 0.0312 &0.0517  &0.0403   &0.0785 \\
&&&0.8898&0.9455&0.9112&0.7833\\
\rowcolor{lightgray}
0.200 & 0.800  &0.60  &0.0334  & 0.0533 & 0.0442&0.0837 \\
& &  & 0.8035 & 0.8850 &0.8320  &0.6783 \\
\rowcolor{lightgray}
0.225& 0.775 &0.55  & 0.0346 &0.0544 &0.0476  & 0.0892\\
&  &  &0.6916  &0.7984 & 0.7321 & 0.5572\\
\rowcolor{lightgray}
0.250& 0.750 &0.50  & 0.0350 &0.0546 & 0.0490 & 0.0561\\
&  &  &0.5554  &0.6851 & 0.6097 &0.3087 \\
\rowcolor{lightgray}
0.275& 0.725 & 0.45 & 0.0350 & 0.0541 &0.0496  & 0.0584\\
&  &  & 0.4090 & 0.5481 & 0.4651 & 0.2064\\
\rowcolor{lightgray}
0.3000&0.7000& 0.40 & 0.0346 &0.0531  & 0.0493 & 0.0599\\
&&  & 0.2671 & 0.3896 & 0.3265 & 0.1203\\
\hline
\hline
\end{tabular}}
 \caption{Type I error rate (upper row) and power (lower row) simulations for the $p$-value and the posterior probability using a sample of size $n=35$, prior parameters $(p,q)=(0.5,0.5), (1,1)$, and $(3,3)$. We use different equivalence margins $\Delta=\theta_2^*-\theta_1^*$.}
    \label{tab:different_margins}
\end{table}
From Table \ref{tab:different_margins} and for small prior parameters, such as $\mathrm{Beta}(0.5,0.5)$, the posterior probability outperforms the $p$-value in terms of the type I error rate control and power for all equivalence margins. This superiority becomes more apparent for smaller equivalence margins compared to wider margins. As the prior parameters increase, the performance of the posterior probability deteriorates slightly compared to the $p$-value. The same trend occurs for large values of the prior parameters, where the posterior probability still performs well compared to the $p$-value for narrow equivalence margins. 

%%%%%%%%%%%%%%%%%%%%%%%%%%%%%%%%%%%%%%%%%%%
\subsection{Power of the FDR Procedure}
\label{appendixB2}
\begin{figure}[ht!]
\begin{center}
\includegraphics[width=17cm,height=14cm]{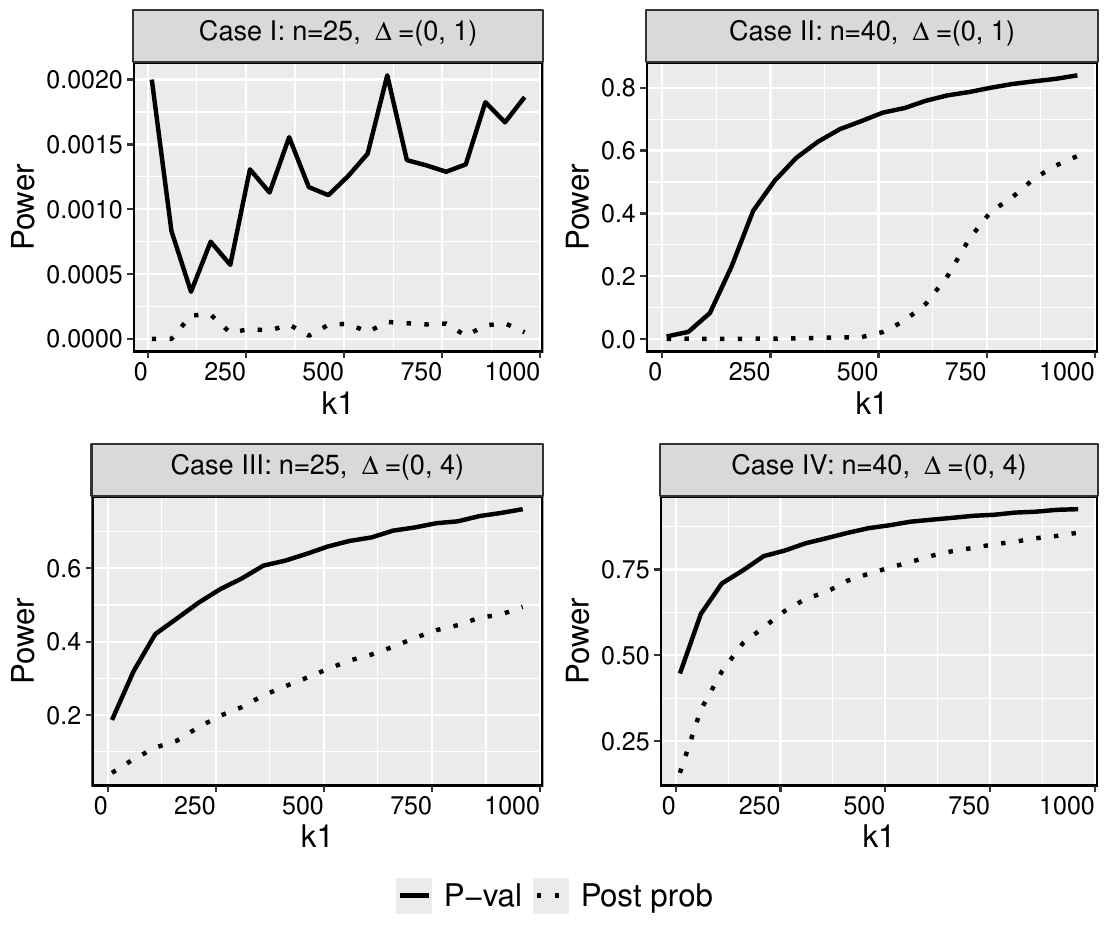}
\caption{An illustration of the effect of different sample sizes and equivalence margins on the monotonicity of the FDR power for the posterior probability in comparison to the $p$-value. We set $\epsilon=0.5$, the number of hypotheses at $k=1{,}000$, and the level of significance at $t=0.05$. Furthermore, we set the number of replications at $r=1{,}000$, the prior standard deviation at $\tau=0.25$, and use different sample sizes and equivalence margins as indicated in the four cases. }
\label{Figure_9_FDR_Power_versus_small_equivalence}
\end{center}
\end{figure}

%%%%%%%%%%%%%%%%%%%%%%%%%%%%%%%%%%%%%%%%%%%
In this Section, we provide additional simulations to investigate the monotonicity of the power of the FDR procedure for small equivalence margins. The simulation steps remain the same as those in Section \ref{simulation}. We set $\epsilon^*=0.5$, the number of hypotheses $k=1{,}000$, and the level of significance at $t=0.05$. Furthermore, we set the number of replications at $r=1{,}000$, the prior standard deviation at $\tau=0.25$, and use different sample sizes and equivalence margins as indicated in the four cases in Figure \ref{Figure_9_FDR_Power_versus_small_equivalence}.

%\begin{lemma}
%\label{prop:monotone_adjustment}
%\end{lemma}

%\begin{proof}
%\end{proof}

%\subsection{Proof of \Cref{cor:cfisher}}

%\section{Proofs}
%\label{sec:proofs_appdx}

%\subsection{Proof of \Cref{thm:adjustment}}
%\label{sec:adjustment_proof}
\end{appendix}
\end{document}